\documentclass[sigconf,singlecolumn]{acmart}

\usepackage{wrapfig}
\usepackage{graphicx}
\graphicspath{{figs/}}
\usepackage[font=small]{caption}
\usepackage[labelformat=simple,font=scriptsize]{subcaption}

\usepackage{algorithm}
\usepackage{algpseudocode}
\usepackage{comment}
\usepackage{relsize}
\usepackage{enumitem}
\usepackage{booktabs}
\usepackage{array}

\newcommand{\eg}{\textit{e.g.},}
\newcommand{\ie}{\textit{i.e.},}

\newcommand{\cat}[1]{\medskip\noindent\textbf{#1.}}
\newcommand{\scat}[1]{\vspace{0.5ex}\noindent\textbf{#1.}}
\newcommand{\fcat}[1]{\noindent\textbf{#1.}}
\newcommand{\sys}{\textsf{Hopper}}

\newcolumntype{C}[1]{>{\centering\arraybackslash}p{#1}}

\renewcommand\footnotetextcopyrightpermission[1]{}

\begin{document}

\title{Predictive Load Balancing for RDMA Traffic}

\author{Erfan Nosrati}
\affiliation{%
  \institution{University of Calgary}
  \city{Calgary}
  \country{Canada}}
\email{erfan.nosrati@ucalgary.ca}

\author{Majid Ghaderi}
\affiliation{%
  \institution{University of Calgary}
  \city{Calgary}
  \country{Canada}}
\email{mghaderi@ucalgary.ca}

\begin{abstract}
Fast training of large machine learning models requires distributed training on AI clusters consisting of thousands of GPUs. The efficiency of distributed training crucially depends on the efficiency of the network interconnecting GPUs in the cluster. These networks are commonly built using RDMA following a Clos-like datacenter topology. To efficiently utilize the network bandwidth, load balancing is employed to distribute traffic across multiple redundant paths. While there exists numerous techniques for load-balancing in traditional datacenters, these are often either optimized for TCP traffic or require specialized network hardware, thus limiting their utility in AI clusters.

This paper presents the design and evaluation of \sys, a new load-balancing technique optimized for RDMA traffic in AI clusters. Operating entirely at the host level, \sys\ requires no specialized hardware or modifications to network switches. It continuously monitors the current path for congestion and dynamically switches traffic to a less congested path when congestion is detected. Furthermore, it incorporates a lightweight mechanism to identify alternative paths and carefully controls the timing of path switching to prevent excessive out-of-order packets.
We evaluated \sys\ using ns-3 simulations and a testbed implementation. Our evaluations show that \sys\ reduces the average and $99$-percentile tail flow completion time by up to $20\%$ and $14\%$, respectively, compared to state-of-the-art host-based load balancing techniques.
\end{abstract}

%


\maketitle

\section{Introduction}
\label{s:intro}

Modern machine learning (ML) models, such as large language models (LLMs) and deep learning recommendation models (DLRMs), contain billions and even trillions of parameters. Fast training of these models requires distributed training on AI clusters consisting of thousands of GPUs~\cite{meta-at-scale}. A critical challenge in distributed training is the communication overhead during GPU synchronization. In large clusters running multiple training jobs, this communication can dominate the training time~\cite{crux-cao}, significantly slowing the process.
The key to faster distributed training is optimizing the cluster network connecting GPU servers to minimize communication overhead. 



Large AI clusters are designed following best practices in high-performance computing (HPC) and datacenter networking. Specifically, they adopt Remote Direct Memory Access (RDMA) for high-speed, low-latency communication~\cite{meta-at-scale}. Modern AI clusters commonly implement RDMA using RoCEv2~\cite{meta-at-scale} due to its compatibility with standard Ethernet. They also utilize hierarchical datacenter topologies (\eg\ leaf-spine~\cite{topo-green}) for flexible scale-out~\cite{HPN}. These topologies provide multiple paths between any two GPU servers for fault tolerance, performance, and scalbility. To maximize network bandwidth utilization and thereby reducing inter-GPU communication time in ML training, it is crucial to implement load balancing in the network to efficiently distribute traffic across different paths~\cite{In-network,MP-RDMA, HPN}. While there is extensive work on load balancing in traditional datacenters~\cite{CONGA,Drill,PLB,Flowbender}, these works generally consider CPU-generated traffic, such as TCP, and thus do not perform efficiently with GPU-generated RDMA traffic~\cite{In-network}. Currently, there is a gap in the literature on practical and deployable load balancing techniques for RDMA in AI clusters, which we aim to address in this work.

A widely used technique for load balancing in today’s datacenters is Equal Cost Multi-Path (ECMP)~\cite{ecmp,hash-xu}. However, numerous studies have shown that ECMP is unable to distribute the load evenly over different paths~\cite{PLB} when flow size distribution is skewed, with a small number of large flows accounting for a disproportionate share of total traffic~\cite{PLB, Flowbender, HeavyKeeper}. This problem is particularly pronounced in distributed ML training workloads, which typically generate fewer but significantly larger flows compared to conventional datacenter workloads, leading to more imbalanced load distribution under ECMP~\cite{meta-at-scale, Strack}.

Several works have been proposed to address the shortcomings of ECMP. For instance, some works use random packet spraying (RPS)~\cite{On-impact-of-PS,Drill} to distribute packets uniformly across all paths, achieving near-perfect load balancing, but introduce a significant number of out-of-order (OOO) packets~\cite{On-impact-of-PS, In-network, MP-RDMA}. 
Other works~\cite{CONGA,let-it-flow} split a flow into flowlets based on inter-packet time gaps in the flow, and send flowlets over different paths. Although flowlet switching is efficient in balancing the load without excessive OOO packets, its efficiency depends on traffic characteristics, \ie\ whether there are flowlets available. This is problematic in AI clusters as hardware-generated RDMA traffic does not include sufficient idle gaps to allow effective flowlet switching~\cite{In-network,MP-RDMA}.
A few works have thus considered switching paths dynamically whenever the current path becomes congested~\cite{Flowbender,MP-RDMA,In-network}. However, these approaches either choose paths to switch to blindly, irrespective of their quality (FlowBender~\cite{Flowbender}), necessitate complex host-side logic to manage multiple paths (MP-RDMA~\cite{MP-RDMA}), or rely on less widely deployed programmable network switches (ConWeave~\cite{In-network}).

In this work, 
we present the design and evaluation of \sys\footnote{At the core of our technique is \textit{hopping} over congested network paths, hence the name \sys.}, a host-based single-path RDMA load balancing technique.
While \sys\ cannot match the performance of hardware-based techniques such as ConWeave, it offers simplicity, practicality, and deployability in current AI clusters. Unlike RPS, \sys\ is robust to changes in path characteristics and topology asymmetries. Moreover, it outperforms other state-of-the-art solutions like FlowBender~\cite{Flowbender} (host-based) and CONGA~\cite{CONGA} (switch-based).
As a host-based solution, \sys\ leverages measured path round-trip-time (RTT) as a signal to detect congested paths and reroute flows to less utilized alternative paths, while minimizing the number of OOO packets during path switching. 
\sys\ requires only standard ECMP support on network switches and minimal host modifications for RTT estimation and path switching. Both of these functionalities have been well-studied by prior work~\cite{Strack, Swift, Flowbender, Timely,PLB, csig,ZTR-RTT}, with a number of efficient solutions already available. 
%

Our work is motivated by FlowBender~\cite{Flowbender}, which targets TCP traffic. Similar to \sys, it performs path switching based on congestion signals (\ie\ ECN marking), but employs a random path selection strategy, which we argue can lead to degraded performance for two key reasons:
\begin{itemize}[leftmargin=*]
\item \textbf{Suboptimal Path Selection:} The newly selected path might also be congested. In such cases, FlowBender continues to use the new congested path for a few RTTs before it decides to switch again, prolonging performance issues. 

\item \textbf{Significant Out-of-Order Packets:} There is a potential for a large difference in path delays between the old and new paths. Thus, with high transmission rates common in RDMA, switching to a random path can cause a significant number of OOO packets. 
\end{itemize}
To avoid these issues, \sys\ employs an \textit{informed} path selection and switching strategy by leveraging the capabilities of modern RDMA NICs (RNICs) for RTT measurement and limited OOO packet handling. First, it only initiates path changes when a clearly better alternative is available, thereby avoiding unnecessary rerouting under heavy network load when all paths are similarly congested. Specifically, when the current path RTT exceeds a certain threshold, \sys\ begins probing for less congested paths by sending small probe packets on two randomly selected alternative paths. 
Once a path is identified as congested, \ie\ its RTT exceeds a second threshold, \sys\ compares the congestion level of the current path with that of two recently probed alternatives. If either probed path offers significantly better conditions, \sys\ changes the path of the flow to the best of the two. Otherwise, the flow continues on the current path.
Second, \sys\ carefully considers the delay difference between the old and new paths. Specifically, packets on the new path are delayed in proportion to the RTT difference between the current and new path, reducing the chance of OOO packets that could trigger retransmissions. Using simulations and testbed experiments, we show that these design optimizations lead to substantial performance gains without introducing any significant overhead, enabled by the capabilities of modern RNICs for packet reordering and RTT measurement.

The contributions of this paper are summarized as follows:
\begin{itemize}
\item We design and evaluate \sys, a host-based load balancing technique for AI clusters that operates at near-RTT granularity, without causing excessive OOO packets. 

\item We develop mechanism for light-weight congestion-aware path selection and switching by leveraging the capabilities of RNICs for RTT measurement and limited packet reordering.

\item We evaluate \sys\ using both ns-3 simulations and a hardware testbed implementation. Our results show that, compared to FlowBender, \sys\ improves average and $99$-percentile flow completion time (FCT) by up to $20\%$ and $14\%$, respectively, for ML training workloads.
\end{itemize}

The rest of the paper is organized as follows. We discuss our motivation for \sys\ in~\S\ref{s:motivation}. Design principles and components of \sys are discussed in~\S\ref{s:design}. Evaluations are presented in~\S\ref{s:eval}. Related works are reviewed in~\S\ref{s:related}, while~\S\ref{s:conc} concludes the paper.

\section{Background and Motivation}
\label{s:motivation}

The primary motivation for our work stems from the observation that random path selection as employed in RPS~\cite{On-impact-of-PS}, FlowBender~\cite{Flowbender}, and PLB~\cite{PLB}, while simple, is not efficient. We argue that with recent advances in RNICs, such as hardware time-stamping and packet reordering, it is now feasible to efficiently implement congestion-aware path selection on the hosts without requiring any support from network switches, modifying packet headers, or degrading the NIC's performance. In this section, we provide a brief overview of RDMA and discuss our motivation in more detail.

\scat{RDMA} 
RDMA is a high-performance communication technology that delivers low latency and high throughput by implementing transport functionalities including congestion control and loss recovery entirely in
NIC hardware. It enables hardware-accelerated direct memory access over the network without involving the CPU or operating system. An RDMA connection is identified by a so-called Queue Pair (QP). The QP is the communication interface that allows user-space applications to send and receive messages via the RDMA transport residing on the RNIC hardware.


\scat{RoCEv2}
RDMA over Converged Ethernet version 2 (RoCEv2) encapsulates RDMA messages within UDP/IP packets, enabling RDMA to operate over standard Ethernet-based datacenter networks. 
RDMA was originally designed to operate over lossless Infiniband fabrics~\cite{infiniband}. To enable RDMA over Ethernet, RoCEv2 relies on Priority Flow Control (PFC)~\cite{pfc} to turn Ethernet into a lossless fabric. However, PFC brings challenges such as congestion spreading and deadlocks~\cite{RDMA-IRN}. To address these issues, several mechanisms are added to RDMA network stack including end-to-end congestion control like DCQCN~\cite{DCQCN}, and selective repeat loss recovery like IRN~\cite{RDMA-IRN}.

\scat{Problem with Out-of-Order Packets}
Modern RNICs support packet reordering in the hardware via IRN. However, to avoid degrading the NIC's performance, it is crucial to limit the number of OOO packets. IRN requires per-connection memory for tracking lost packets and rearranging OOO packets. Commercial RNICs have limited on-chip SRAM memory~\cite{SRNIC}. For example, NVIDIA CX-5 RNIC has only about $2$~MB on-chip memory~\cite{rpc-kalia}. Allocating even one BDP (bandwidth-delay product)-sized reordering buffer consumes $0.1$~MB memory (network bandwidth $100$~Gbps and RTT $8$~$\mu$s) per connection, severely limiting the number of connections the RNIC can support.


\scat{Problem with Random Packet Spraying}
RPS looks like the perfect load balancing approach as it uniformly distributes the load across all paths. But, that is exactly where the problem lies. Not only this creates a massive number of OOO packets~\cite{In-network}, overwhelming the RNIC resources, but also is vulnerable to network asymmetries. For uniform load balancing, RPS requires a symmetric network topology with consistent path characteristics.
However, in large AI clusters, device failures (\eg\ links and switches) are common~\cite{HPN}, leading to an asymmetric topology. Moreover, transient congestion due to rerouted traffic can also lead to asymmetries in path characteristics. With such asymmetries, uniformly sending packets over all paths leads to inflated tails for flow completion times. This degradation is particularly detrimental for distributed ML training, where training progress is gated by the completion time of the slowest flow in a collective communication operation~\cite{Packet-trimming}.

\scat{Summary} 
An ideal solution is to make RPS consider path congestion on a packet-by-packet basis. But the overhead of such an approach is prohibitive. A more practical solution is to increase the load balancing granularity to amortize the path assessment overhead. This is exactly what \sys\ aims to achieve by operating at RTT granularity, slower than per-packet but faster than per-flow load balancing. While \sys\ can't achieve perfect uniform load distribution as RPS, it does not create as many OOO packets and is robust against topology changes.

\section{Design}
\label{s:design}

%
Our design for \sys\ leverages the ability of modern RNICs~\cite{nvidia_connectx5, nvidia_connectx6} for limited packet reordering and precise time-stamping. In the following subsections, we present the design of \sys\ and discuss its various components. An example of \sys's workflow is presented in Appendix~\S~\ref{a:workflow}.



			\begin{algorithm}[t]
			\caption{\textbf{Alg.~1} \sys\ Control Logic in Each RTT Epoch.}
			\label{a:hopper}
			\begin{algorithmic}
				\State \texttt{avg\_rtt} $\gets 0$, 
				 \texttt{probe} $\gets$ \texttt{true},
				 \texttt{switch} $\gets$ \texttt{true}
				
				\For{every \texttt{new\_rtt} measurement} 
				\State $\texttt{avg\_rtt} \gets \alpha\cdot \texttt{new\_rtt}
					+ (1-\alpha)\cdot \texttt{avg\_rtt}$
				\If{\texttt{avg\_rtt} $>$ \texttt{th\_probe \& probe == true}}
					\State \texttt{alt\_paths} $\gets$ \Call{ProbePaths}{ }
					\State \texttt{probe $\gets$ false}
				\EndIf
				\If{\texttt{avg\_rtt} > \texttt{th\_cong \& switch == true}}
					\State \Call{SwitchPath}{\texttt{alt\_paths}}
					\State \texttt{switch $\gets$ false}
				\EndIf
				\EndFor
			\end{algorithmic}
			\end{algorithm}

\cat{Design Overview}
\sys\ operates over control epochs, where the duration of each epoch is set to one RTT. It consists of three main modules: (1) a congestion detection module, (2) a path probing module, and (3) a path switching module. Together, these modules enable \sys\ to dynamically reroute flows to less congested paths by only manipulating packet headers on the host to affect standard ECMP hashing decisions. The high-level pseudo-code for \sys\ is presented in Alg.~\ref{a:hopper}. Below, we describe the functionality of each module.

\subsection{Congestion Detection Module}
\fcat{Congestion Signal}
The two most widely used congestion signals are Explicit Congestion Notification (ECN) and Round-Trip-Time (RTT). These signals are often readily available at RDMA transport layer, but could also be obtained independently with low overhead~\cite{csig}. For example, DCQCN~\cite{DCQCN} relies on ECN marking by switches, while Timely~\cite{Timely} uses RTT measurements to infer congestion. 
\sys\ leverages RTT for path-level congestion detection, as it is supported by commodity NICs~\cite{Swift}, requires no switch support, and is particularly suited for path probing. 
Since probing typically involves transmitting only a small number of packets, ECN may not reliably indicate congestion due to insufficient marking opportunities. 
%
On the other hand, ECN is capable of signaling emerging congestion earlier, even before RTT increases become apparent~\cite{Strack}. While \sys\ can be further optimized to leverage ECN markings to initiate path probing proactively, our current implementation is based on RTT measurements. 


\scat{Congestion Detection}
\sys\ monitors measured per-packet RTT (denoted by \texttt{new\_rtt} in Alg.~\ref{a:hopper}) on the sending host, \eg\ using ACKs in IRN~\cite{RDMA-IRN}. It keeps a moving average of measured RTTs over a control epoch of one RTT. The current path is considered congested when the average RTT exceeds a predefined threshold \texttt{th\_cong}. 
The threshold \texttt{th\_cong} can be tuned to best match the specific network architecture and workload characteristics. Also, the moving average parameter $\alpha$ can be increased (decreased) to make \sys\ more (less) responsive to RTT increases.

%
%

\subsection{Path Probing Module}
\fcat{Probe Mechanism} 
Given modern RNICs’ support for limited out-of-order delivery, probing can be integrated directly into data transmission, sending a few data packets along alternate paths. 
Therefore, in our current design, \sys\ uses out-of-band probing packets to scan alternative paths. However, sending probing packets, if not carefully controlled, can increase network load, especially when the network is already congested. Moreover, probing a new path in RDMA requires using a new QP, which could affect the scalability of RNICs~\cite{SRNIC}, if not controlled carefully.

\scat{Probe Initiation}
To address these challenges, \sys\ employs the power-of-two-choices strategy. If the average RTT exceeds a threshold \texttt{th\_probe}, probing for alternate paths starts. To this end, \sys\ probes two randomly selected unexplored paths by creating two QPs bound to distinct UDP source ports. It records these source ports along with the corresponding measured delays to inform rerouting decisions. In our testbed implementation, we perform profiling to map specific source ports to distinct paths, ensuring consistent path selection during probing. 
To prevent redundant probing, \sys\ employs timestamping to track recently explored paths and avoids selecting any path again if it was probed within the last \texttt{ttl\_probe} interval. 

\subsection{Path Switching Module}
\fcat{Path Selection} 
\sys\ switches the current path only if it is congested and there is a less congested path available. Otherwise, it will wait for one RTT epoch to initiate probing again. 
To prevent rerouting to paths exhibiting similar congestion levels, \sys\ initiates rerouting only when the alternative path demonstrates a substantially lower RTT than the current path, exceeding a configurable margin $\delta_{rtt}$. This parameter can be optimized to balance performance and stability by avoiding path switches that do not result in meaningful improvements. By preventing unnecessary path switches, this approach would also reduces the number of out-of-order packets.

\scat{Path Switching}
To switch to a new path, \sys\ simply starts using the corresponding QP that was used to probe the path and releases the QP of the old path. If neither path offers improvement, \sys\ frees the probed QPs but retains the source port and delay information for a few RTTs to avoid repeatedly probing the same congested paths. 
Although this approach may introduce some delay in switching away from congested paths, it effectively prevents excessive path switching when all links are persistently congested. Additionally, it avoids the issue where multiple flows from congested paths are rerouted to the same alternative path due to aggressive probing frequency. 

\begin{figure}[t]
		\centering
		\includegraphics[width=\linewidth]{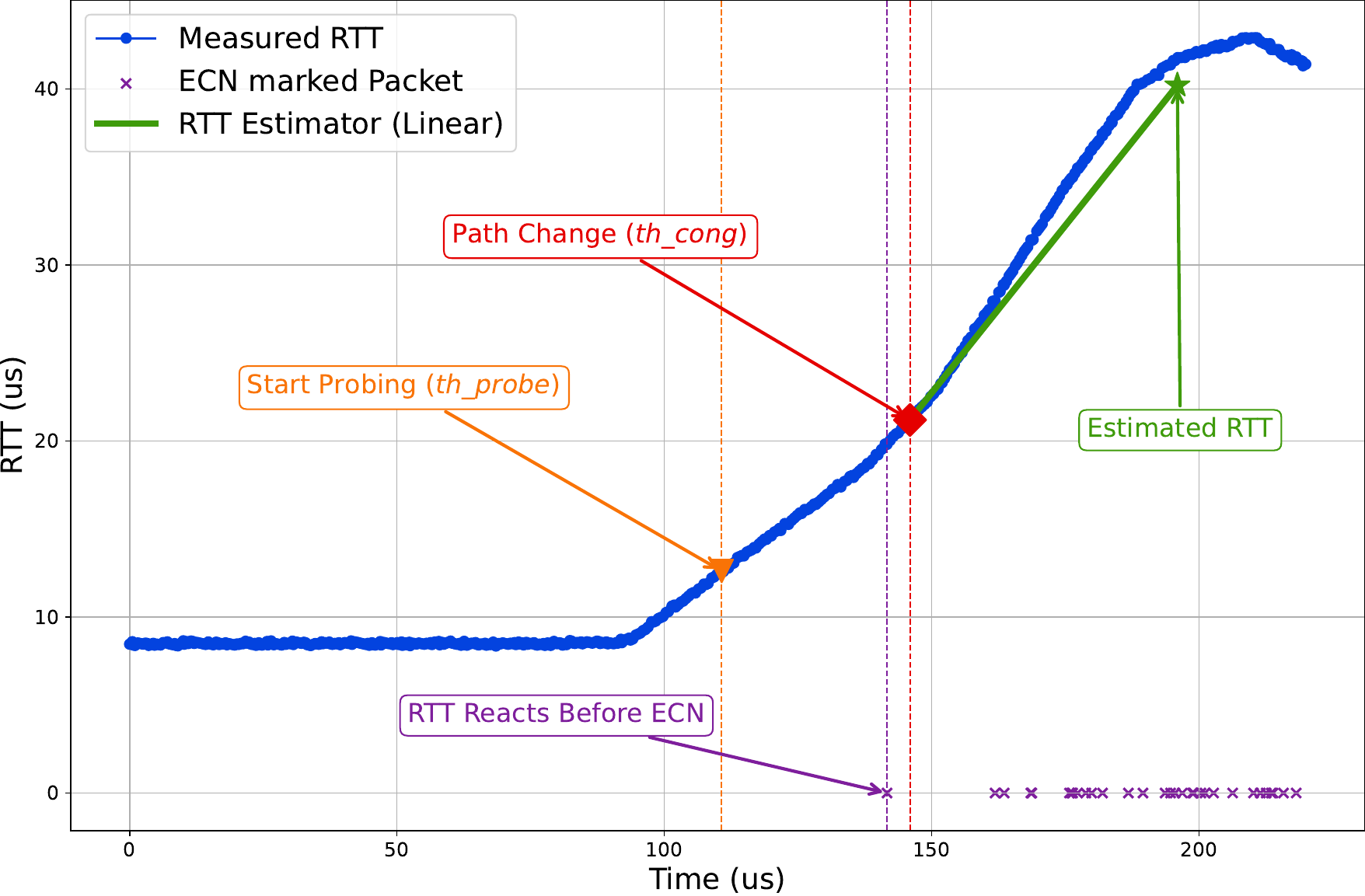}
		\caption{\sys\ employs a linear estimator to predict path RTT.}
		\label{fig:linear_est}
\end{figure}
\scat{Reducing Out-of-Order Packets}
Since the RTTs of both the current and alternative paths are available, \sys\ can delay rerouting by a duration proportional to the RTT difference between the two paths. This delay helps reduce out-of-order packet delivery at the receiver~\cite{MP-RDMA}. However, if not carefully implemented, such delayed rerouting can introduce unnecessary latency, contradicting RDMA’s low-latency goals and degrading end-to-end performance. 
To calculate a suitable delay, \sys\ needs to estimate RTT on the current path for the in-flight packets. As depicted in Fig.~\ref{fig:linear_est}, we employ a linear regression model for this purpose. The model calculates the rate of RTT increase using \texttt{new\_rtt} measurements in the current RTT epoch and extrapolates this trend based on the number of in-flight packets to estimate the RTT of the last packet on the congested path. Although the estimated RTT may overestimate the actual path delay, since RTT increases tend to stabilize once queues stop growing, \sys\ uses this value as a conservative upper bound. 

\section{Evaluation}
\label{s:eval}

We use ns-3~\cite{ns3_website} simulations and a tested implementation to evaluate \sys\ under traditional datacenter and ML workloads. 
The objective of our evaluation is to address the following questions:
\begin{itemize}
\item \textit{How much improvement congestion-aware path selection can achieve compared to random path selection?} To answer this question, we compare \sys\ with FlowBender~\cite{Flowbender}.

\item \textit{How effective dynamic load-balancing is compared to flowlet switching for RDMA?} To answer this question, we compare \sys\ with CONGA~\cite{CONGA}.

\item \textit{What is the performance gap between host-based and in-network load balancing?} To answer this question, we compare with ConWeave~\cite{In-network}.
\end{itemize}

\subsection{Simulation Experiments}

\begin{table}[t]
	\centering
	\caption{\sys\ parameters. \vspace{-2ex}}
	\begin{tabular}{cc}
		\toprule
		\textbf{Parameter} & \textbf{Value} \\
		\midrule
		$\alpha$   & 1 \\
		\texttt{th\_prob} & 1.5 × base RTT \\
		\texttt{th\_cong} & 2.5 × base RTT \\
		\texttt{ttl\_prob} &  4.0 × base RTT \\
		$\delta_{rtt}$   & 80\% \\
		\bottomrule
	\end{tabular}
	\label{tab:parameters}
\end{table}
We build on the ns-3 implementation provided by ConWeave, using the same network topology, transport parameters, and workload generator. However, in addition to datacenter workloads used in ConWeave, we also incorporate an ML training workload to specifically study the behavior of \sys\ and other techniques in AI clusters. The parameters used for \sys\ in the simulations are shown in Table~\ref{tab:parameters}. These values are found to work well in our setup.

\begin{figure}[H]
		\centering
		\includegraphics[width=\linewidth]{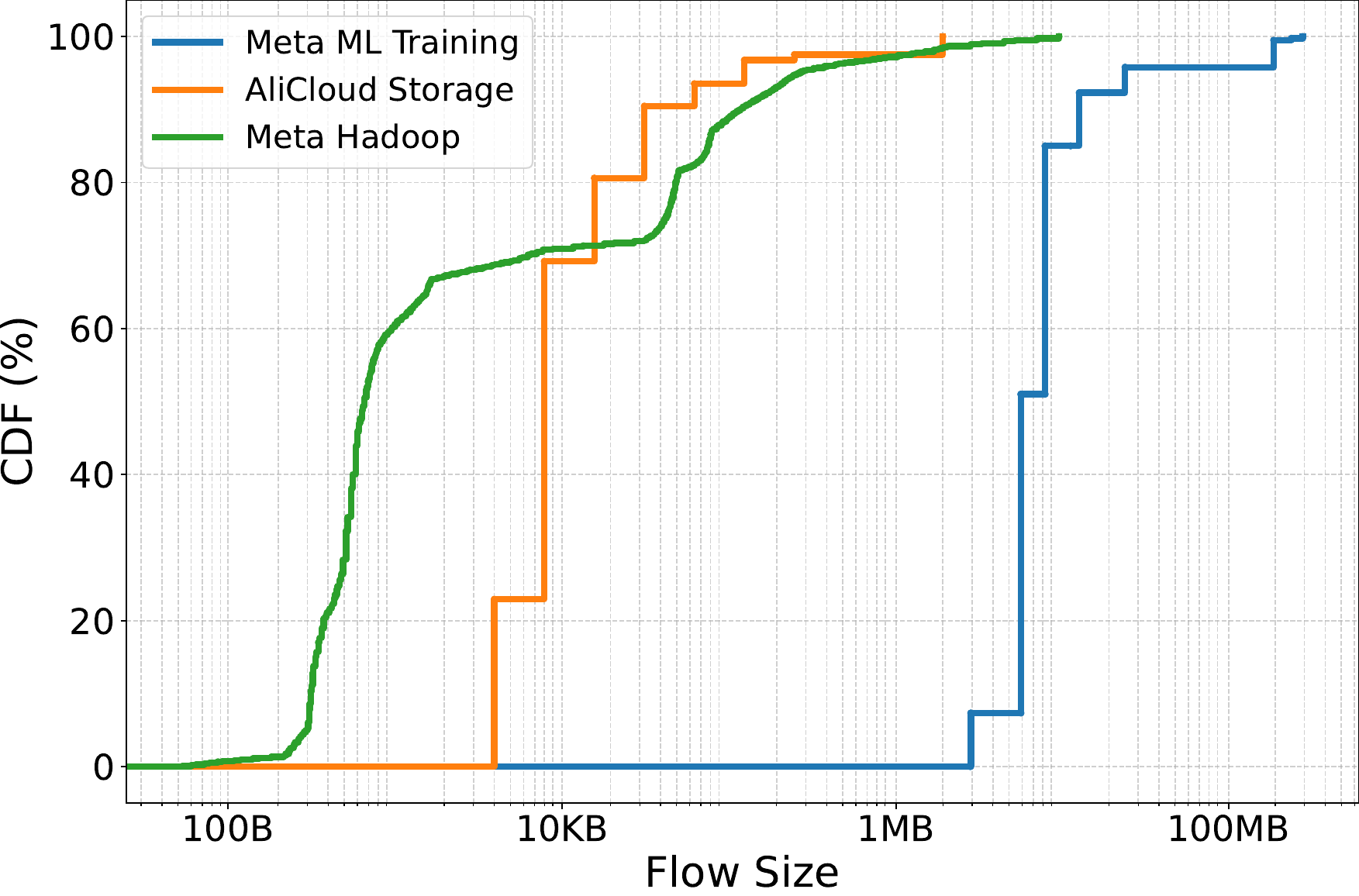}
		\caption{Traffic characteristics of representative workloads. The flow sizes in ML Training workload are substantially larger than those in datacenter workloads.}
		\label{fig:cdf}
\end{figure}

\smallskip
\subsubsection{Setup and Methodology}~\\
\fcat{Network Topology} 
We adopt a leaf-spine topology comprising 128 servers and 16 switches. Each group of 16 servers connects to a dedicated leaf switch, and all 8 leaf switches are fully connected to 8 spine switches. All links are 100 Gbps with a latency of $1~\mu s$, for a base RTT of $8~\mu s$.


\begin{figure*}[t]
	\centering
	\begin{subfigure}[t]{0.24\textwidth}
		\centering
		\includegraphics[width=\textwidth]{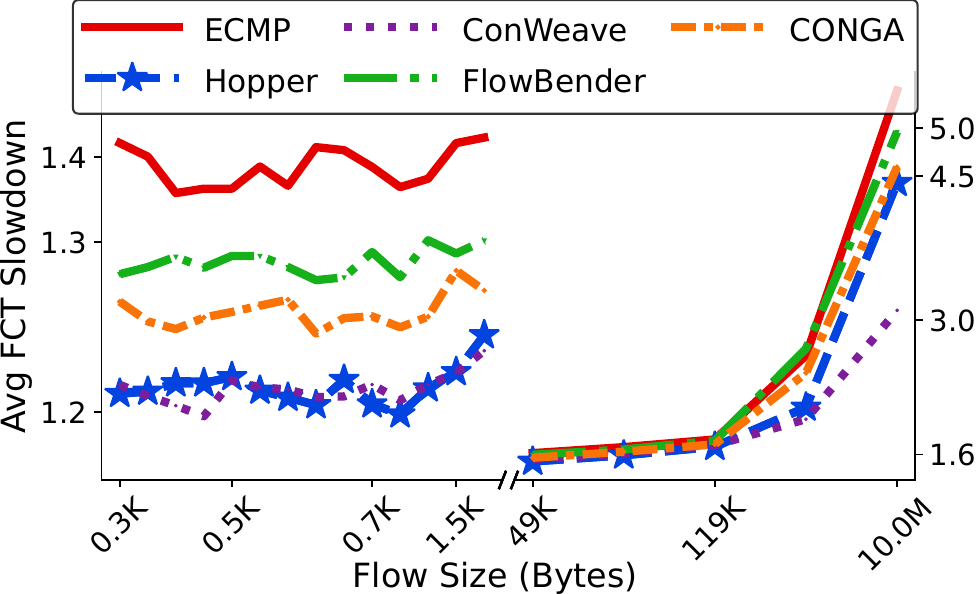}
		\caption{\scriptsize $50\%$ Load (Avg)}
		\label{fig:MetaHadoop:avg1}
	\end{subfigure}
	\hfill
	\begin{subfigure}[t]{0.24\textwidth}
		\centering
		\includegraphics[width=\textwidth]{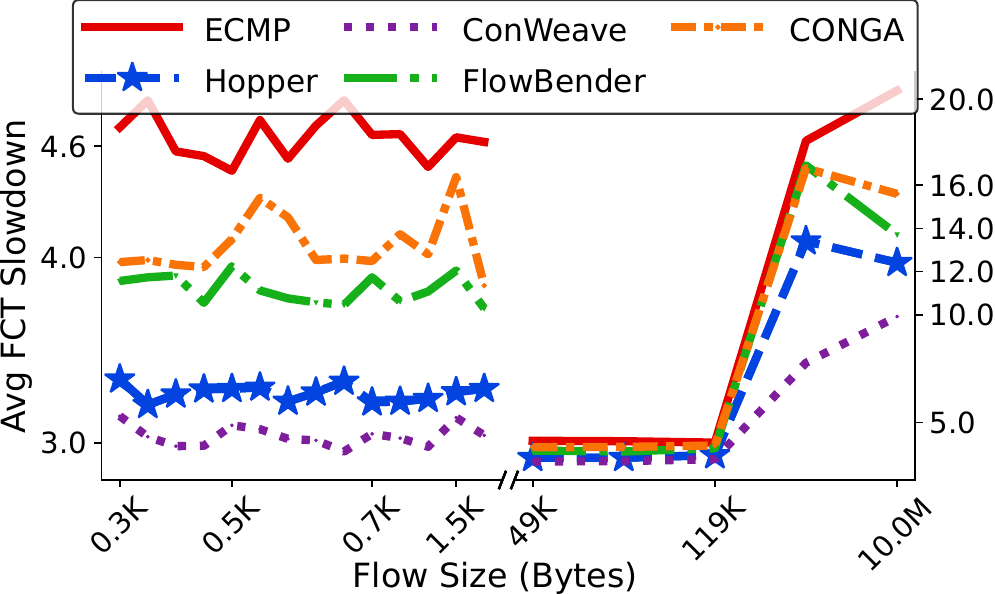}
		\caption{\scriptsize $50\%$ Load (p$99$)}
		\label{fig:MetaHadoop:p991}
	\end{subfigure}
	\hfill
	\begin{subfigure}[t]{0.24\textwidth}
		\centering
		\includegraphics[width=\textwidth]{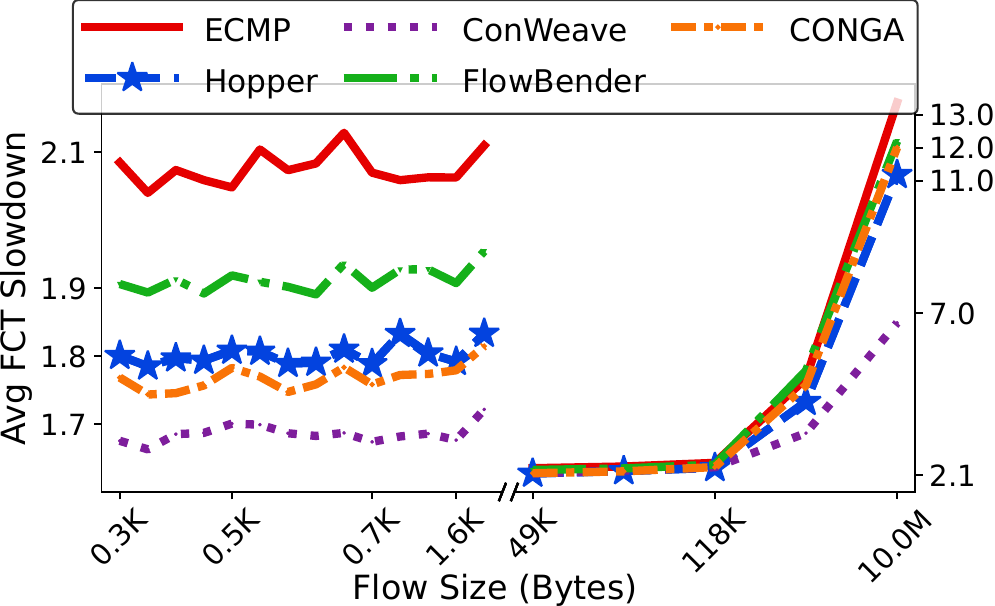}
		\caption{\scriptsize $80\%$ Load (Avg)}
		\label{fig:MetaHadoop:avg2}
	\end{subfigure}
	\hfill
	\begin{subfigure}[t]{0.24\textwidth}
		\centering
		\includegraphics[width=\textwidth]{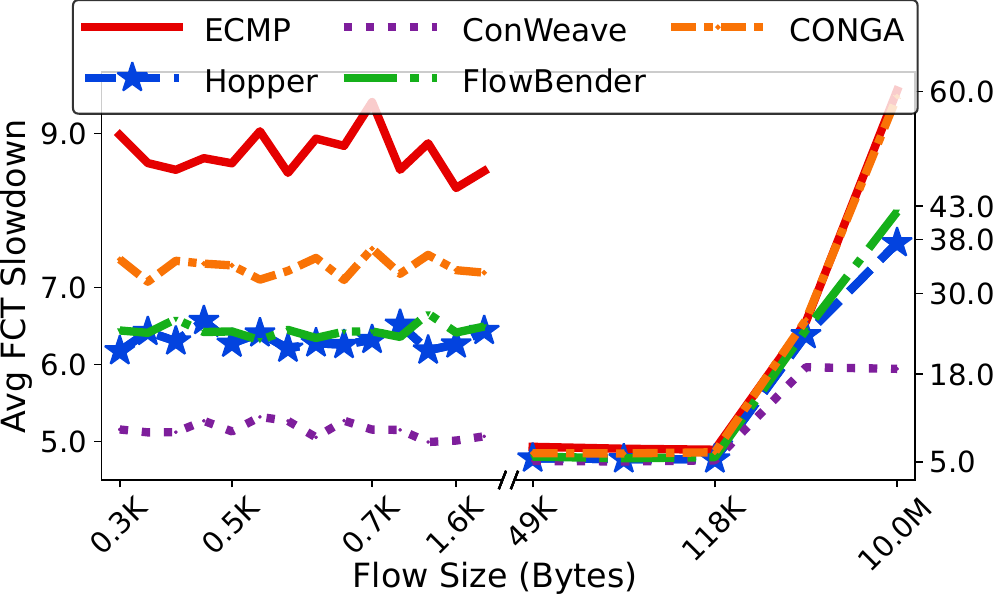}
		\caption{\scriptsize $80\%$ Load (p$99$)}
		\label{fig:MetaHadoop:p992}
	\end{subfigure}
	\caption{Average and tail FCT slowdown for Hadoop workload at $50\%$ and $80\%$ average network load.}
	\label{fig:hadoop:slowdowns}
\end{figure*}

\begin{figure*}[t]
	\centering
	\begin{minipage}[t]{0.49\textwidth}
		\centering
		\begin{subfigure}[t]{0.49\textwidth}
			\centering
			\includegraphics[width=\textwidth]{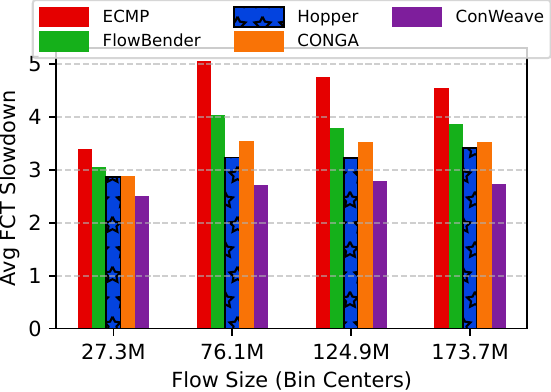}
			\caption{\scriptsize 50\% Load (Avg)}
			\label{f:smart:avg40}
		\end{subfigure}%
		\hfill
		\begin{subfigure}[t]{0.49\textwidth}
			\centering
			\includegraphics[width=\textwidth]{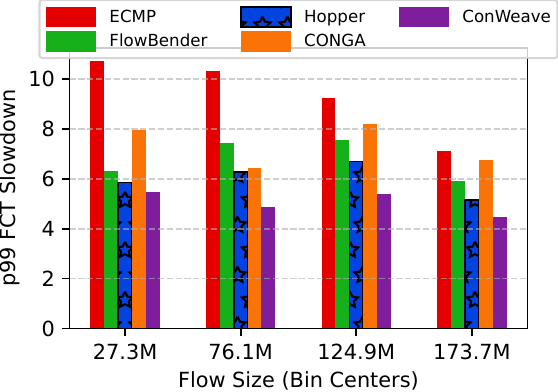}
		\caption{\scriptsize 50\% Load (p$99$)}
			\label{f:smart:p9940}
		\end{subfigure}
	\end{minipage}
	\hfill
	\begin{minipage}[t]{0.49\textwidth}
		\centering
		\begin{subfigure}[t]{0.49\textwidth}
			\centering
			\includegraphics[width=\textwidth]{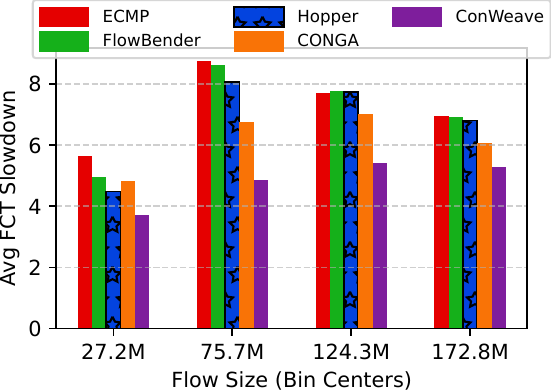}
			\caption{\scriptsize 80\% Load (Avg)}
			\label{f:smart:avg50}
		\end{subfigure}%
		\hfill
		\begin{subfigure}[t]{0.49\textwidth}
			\centering
			\includegraphics[width=\textwidth]{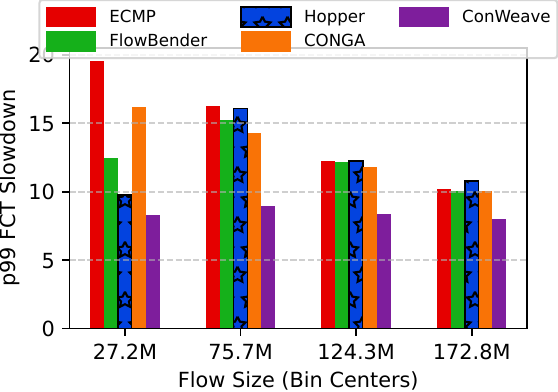}
		\caption{\scriptsize 80\% Load (p$99$)}
			\label{f:smart:p9950}
		\end{subfigure}
	\end{minipage}
	\caption{Average and tail FCT slowdown for ML Training workload at $50\%$ and $80\%$ average network load.}
	\label{f:ml:slowdown}
\end{figure*}

\scat{Transport}
We use DCQCN~\cite{DCQCN} for congestion control, which matches the algorithm used by NVIDIA CX-5 RNICs in our testbed. We set DCQCN parameters identical to those used in ConWeave. We also implement and extended version of IRN~\cite{RDMA-IRN} for packet reordering on RNICs. When an OOO packet is received, it is buffered and acknowledged normally as long as its sequence number falls within a predefined threshold (set to $30$ packets) relative to the expected sequence number. If the sequence number exceeds this threshold, the receiver initiates loss recovery by sending a NACK containing both a cumulative acknowledgment and a selective acknowledgment (SACK). All other aspects of IRN, including limiting in-flight data to one BDP, remain unchanged.
    
\scat{Workloads} 
We conduct experiments using traditional datacenter workloads, including AliCloud Storage~\cite{HPCC} and Meta Hadoop~\cite{Meta-hadoop}, as well as traffic patterns representing those of distributed ML training. Specifically, we utilize the collective message size distribution reported in~\cite{meta-at-scale}, which analyzes ML training jobs of up to $128$ GPUs on Meta’s internal AI cluster. The workload includes a mix of collective communication operations~\cite{nccl_website}, with AllReduce being common in DDP~\cite{ddp}, while AllGather and ReduceScatter being prevalent in FSDP~\cite{fsdp}. Message sizes vary depending on the model architecture and training strategy, capturing a broad range of realistic communication demands. The CDFs of these workloads are shown in Fig.~\ref{fig:cdf}. We used the ConWeave's load generator to simulate network traffic by randomly selecting a sender–receiver pair and assigning a flow size based on the workload distribution. Flow start times follow a Poisson distribution and are chosen to create two network load scenarios, a moderate load scenario at $50\%$ and a high load scenario at $80\%$ network load.


\scat{Baseline Schemes}
We compare \sys\ with FlowBender, CONGA, and ConWeave. We do not include PLB in our evaluation, as it is specifically designed for TCP flows and behaves similar to ECMP in RDMA. Also, like FlowBender, it employs a random path selection strategy. 

\scat{Performance Metric}
We use flow completion time (FCT) slowdown as the primary performance metric, defined as the ratio of a flow’s actual completion time to its baseline completion time in an unloaded network. We present both the average and $99$-percentile tail (p$99$) of the slowdown.

\smallskip
\subsubsection{Results and Discussion}~\\
\label{result_and_discussion}
\fcat{Datacenter Workloads}
Fig.~\ref{fig:hadoop:slowdowns} shows the results for Hadoop workload. The results for AliCloud workload are similar and included in the Appendix~\ref{s:alibaba}. From the workload distribution in Fig.~\ref{fig:cdf}, we observe that only a small percentage of flows are between $2$KB and $49$KB. Thus, for the sake of presentation, the $x$-axis in the plots is broken into two regions, the left region shows short flows ($<2$KB), while the right region shows large flows ($>49$KB). The largest flow size in this workload is $20$MB, and only less than $5\%$ of flows are larger than $266$KB. Therefore, we caution that the reported results for $10$MB flow size are not stable. Instead, for large flow sizes, we focus on the results reported for the ML Training workload in the next subsection. The main observation is that \sys\ consistently outperforms FlowBender for both average (by up to $7.8\%$) and p$99$ tail slowdown (by up to $19.6\%$) across different loads and flow sizes. Interestingly, even though both \sys\ and FlowBender operate at RTT granularity, and thus do not load balance short flows, we see significant improvement compared to ECMP due to them load balancing large flows more effectively. 
We also observe that \sys\ similarly outperforms CONGA except for the average slowdown in high load, where they are both comparable. Finally, in moderate load, \sys's performance is comparable to ConWeave for small flows, with a gap of less than $8.3\%$.


\scat{ML Workload}
For the ML Training workload, traffic is more concentrated around specific flow sizes. Thus, for presentation clarity, as depicted in Fig.~\ref{f:ml:slowdown}, we divide flows into four bins to show FCT slowdown per size category. Recall that this workload mainly consists of large flows that represent collective communication operations. Therefore, not surprisingly, the behavior of different approaches is consistent with the scenario of large flows in datacenter workloads. Specifically, we observe that 1) \sys\ outperforms FlowBender across all scenarios, achieving up to $20\%$ and $14\%$ improvement in average and p$99$ FCT slowdown, respectively, and 2) when the network is moderately loaded, \sys\ even outperforms CONGA by up to $10\%$ and $23\%$ in average and p$99$ FCT slowdown, respectively. However, with high network load, switch-based techniques generally outperform host-based ones, specially for the larger flow sizes.

\subsection{Testbed Experiments}
\label{eval:testbed}
We implement a software-based prototype of \sys\ on a hardware testbed, as depicted in Fig.~\ref{fig:testbed-topology}. The hardware components of the testbed include Dell PowerEdge R740 servers equipped with NVIDIA CX-5 RNICs~\cite{nvidia_connectx5}, and a Dell PowerSwitch S5248F-ON switch running SONiC~OS~\cite{sonic2025}.

\smallskip
\subsubsection{Setup and Methodology}~\\
\fcat{Implementation}
Due to the black-box nature of CX-5 RNICs, implementing \sys\ directly within the RNIC hardware is not feasible. While some prior works~\cite{PACC,BiCC,LSCC} implement similar mechanisms using software frameworks such as DPDK~\cite{dpdk_website}, we choose to preserve compatibility with the RDMA communication model. To that end, the control logic of \sys\ is implemented at user space leveraging the Linux \texttt{rdma-core} library~\cite{rdma-core}, while the actual data transfers are implemented using standard RDMA verbs. 
To send a flow, we divide it into chunks and send chunks via RDMA. The chunk size specifies the granularity at which \sys\ can switch paths. We report experiment results with two chunk sizes for data transfers, namely $10$MB and $1$MB. We find that chunk sizes less than $100$KB degrade performance due to increased frequency of system calls and context switches. For probing alternative paths, we use a $10$KB chunk size to reduce probing overhead. To measure RTT on a path, we measure the time between initiating an RDMA send operation and receiving the corresponding completion event in the completion queue. 

\begin{figure}[t]
	\centering
	\includegraphics[width=0.95\linewidth]{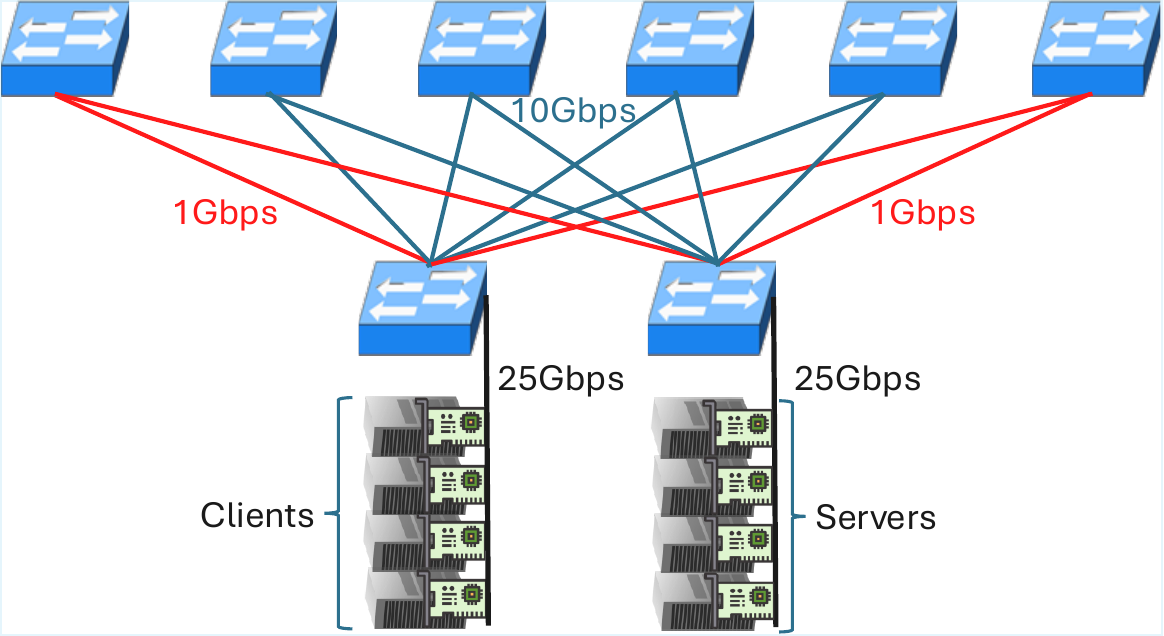}
	\caption{The leaf-spine topology of the testbed with asymmetric path bandwidths.}
	\label{fig:testbed-topology}
\end{figure}

\scat{Network Topology}
Our physical testbed has a leaf-spine topology (see Fig.~\ref{fig:testbed-topology}) with two leaf switches and six spine switches, created on our physical switch using VRF. 
Each RNIC is equipped with two 25~Gbps ports, each assigned to one host. 
Each leaf is connected to four hosts in a rack. Also, each leaf is connected to four spine switches via 10~Gbps links and to the remaining two spine switches via 1~Gbps links, creating deliberate asymmetry in path bandwidths.

\scat{Transport}
All experiments use DCQCN as the congestion control algorithm, configured with the default parameters recommended by NVIDIA for CX-5 RNICs~\cite{nvidia_connectx5}.

\scat{Workload}
We evaluate \sys\ using an AllReduce collective communication pattern, where each sender transmits data to its corresponding receiver. The size and frequency of these collective operations are derived from the GPT-3 model training traces~\cite{Astrasim2}. We extract the communication sizes from the traces and generate flow workloads using a custom script. The workload consists of 204 flows, and we repeat each experiment five times, reporting the average to mitigate the effects of randomness. 
ML training consists of several rounds. In each round, nodes synchronize their intermediate computations using AllReduce before starting a new round.
In our testbed, to simplify implementation, at the end of each round of training, clients send a completion message to a centralized server. The next round begins only after the server confirms that all clients have completed the previous round, \ie\ have completed their associated flows. 

\scat{Baseline Schemes}
We compare \sys\ against two baselines, namely ECMP and a simplified software version of FlowBender. In our FlowBender implementation, congestion is inferred from RTT increases instead of ECN markings. As mentioned earlier, ConWeave and CONGA both require switch modifications, ConWeave in particular requires P4 switches, which we do not have in our testbed. Thus, our testbed comparison is limited to ECMP and FlowBender.

\scat{Performance Metrics} 
We use the following performance metrics for comparison: 1) the average and tail FCT, 2) the average total training time, and 3) the average link utilization. Specifically, we examine the utilization of 1G links, as their use indicates poor load balancing. Since transmitting flows from all four clients over a single 10G link offers higher throughput than on 1G links, higher use of 1G links reflects inefficient path selection by the load balancing technique.

\begin{figure*}[t]
	\begin{subfigure}[b]{0.73\textwidth}
		\includegraphics[width=0.9\linewidth,height=3cm]{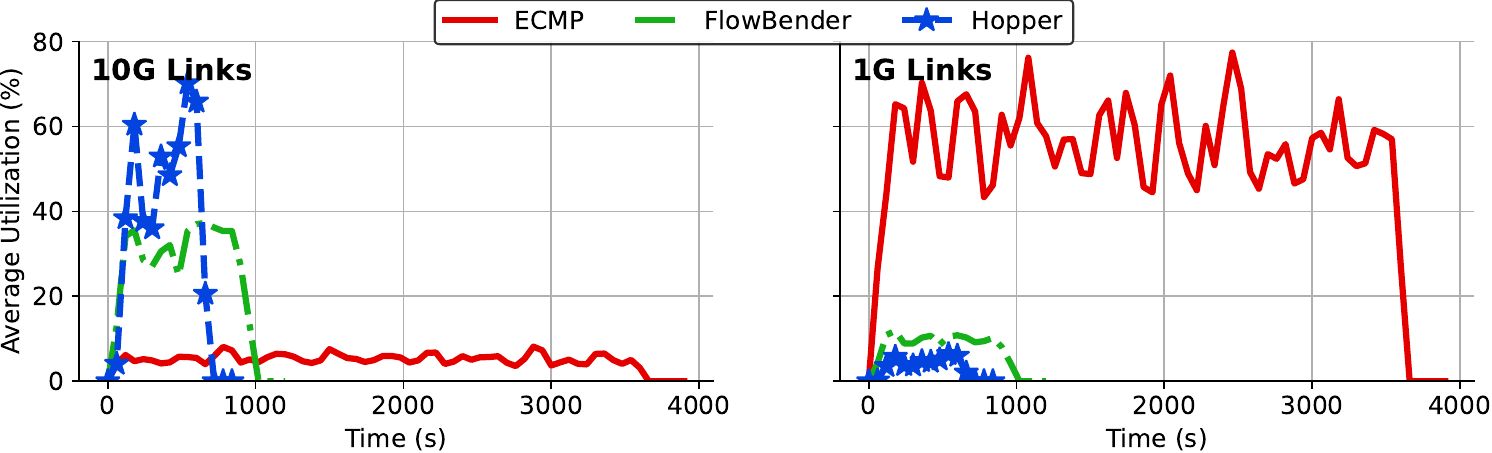} 
		\vspace{-1ex}
		\caption{\scriptsize Link utilization (1MB)}
		\label{fig:util_1mb}
	\end{subfigure}
	\begin{subfigure}[b]{0.23\textwidth}
		\includegraphics[width=\linewidth]{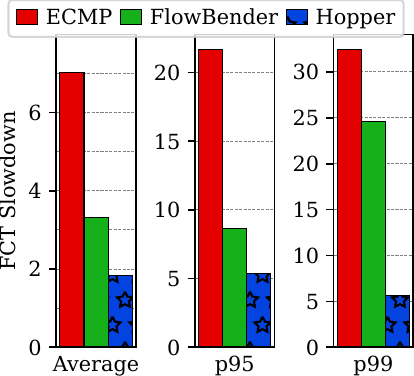}
		\caption{\scriptsize FCT Slowdown (1MB)}
		\label{fig:fct_1mb}
	\end{subfigure}
	\begin{subfigure}[b]{0.73\textwidth}
		\vspace{2ex}
		\includegraphics[width=0.9\linewidth,height=3cm]{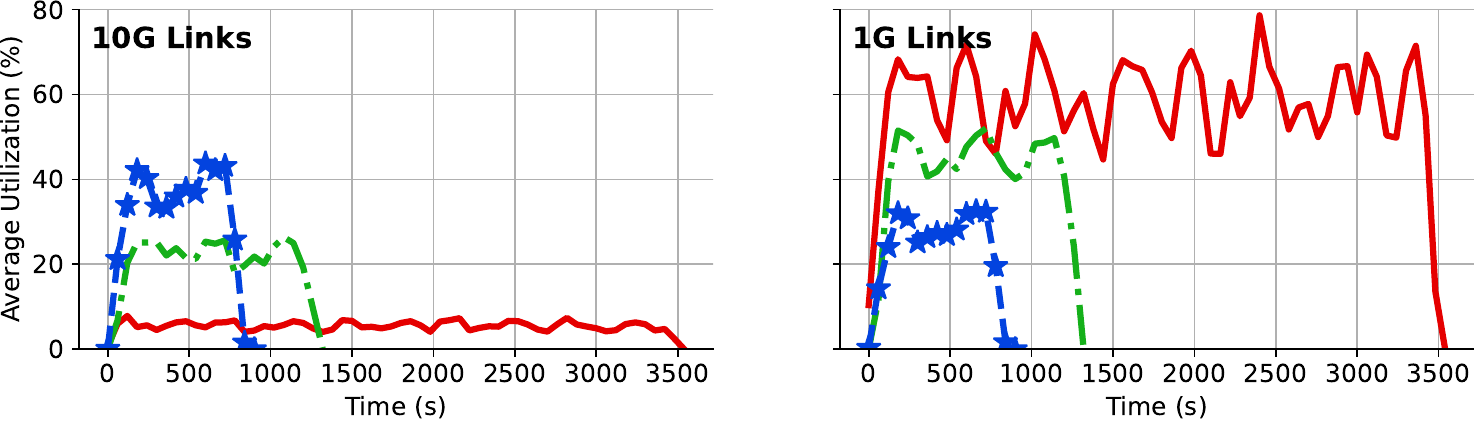} 
		\vspace{-1ex}
		\caption{\scriptsize Link utilization (10MB)}
		\label{fig:util_10mb}
	\end{subfigure}
	\begin{subfigure}[b]{0.23\textwidth}
		\includegraphics[width=\linewidth]{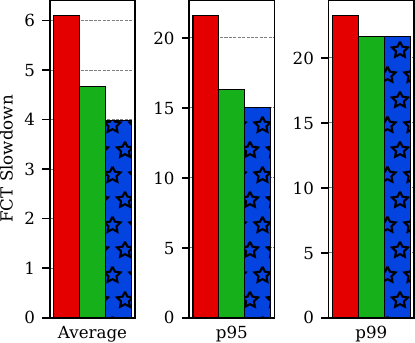}
		\caption{\scriptsize FCT Slowdown (10MB)}
		\label{fig:fct_10mb}
		\vspace{1ex}
	\end{subfigure}
	\caption{Performance of different load balancing approaches in the testbed with $1$MB and $10$MB chunks.}
	\label{fig:util_vs_fct_combined}
\end{figure*}

\smallskip
\subsubsection{Results and Discussion}~\\
\fcat{Avoiding Congested Paths}
When a flow starts, it is initially assigned to a random path, which potentially could be a 1G path. However, both FlowBender and \sys\ will try to switch the flow to other paths after a short time. Figs.~\ref{fig:util_1mb} and~\ref{fig:util_10mb} show the average utilization of 1G and 10G links in the network when using different chunk sizes. We make the following observations. First, \sys\ consistently utilizes 10G links more than FlowBender, by up to $35\%$, clearly demonstrating its ability to discover less congested paths. Second, as the chunk size increases so does the 1G link utilization under both \sys\ and FlowBender. This is because whenever these algorithms choose a 1G path, they stay on that path for a longer time when using large chunks. Yet, the performance degradation of FlowBender is more pronounced, almost twice that of \sys, due to it having a higher chance of choosing 1G paths, as it employs a random path selection strategy.

\scat{FCT Slowdown}
As shown in Figs.~\ref{fig:fct_1mb} and~\ref{fig:fct_10mb}, \sys\ outperforms FlowBender in all scenarios for both average and tail slowdown, except for p$99$ with $10$MB chunks, where it matches FlowBender. As with link utilization, the penalty of choosing a congested path with $10$MB chunk is higher for both approaches, thus leading to a smaller gap between their performances. We note that, smaller chunk sizes result in more granular load balancing decisions, but also higher system overhead in our implementation. For example, with ECMP, we see from the figures that its slowdown increases for $1$MB chunks, as it does not benefit from smaller chunk sizes. For both \sys\ and FlowBender, the chunk size provides a trade-off between load balancing granularity and system overhead. In our testbed, $1$MB chunk size provides a good trade-off. Specifically, with $1$MB chunks, \sys\ improves the average, p$95$, and p$99$ FCT slowdown by $45\%$, $61.9\%$, and $77.2\%$, respectively, compared to FlowBender.
%
%

\scat{Training Time} 
Figs.~\ref{fig:util_1mb} and~\ref{fig:util_10mb} also show the average total training time (multiple training rounds) under different load balancers. Specifically, with $1$MB chunks, the average training time is $1080$ and $715$ seconds under \sys\ and FlowBender, representing $51\%$ reduction in training time. Even with $10$MB chunks, the average training time is $1320$ and $890$ seconds under \sys\ and FlowBender, representing $48.3\%$ reduction in training time. This clearly demonstrates the superiority of \sys\ for ML training workloads.


\section{Related Work}
\label{s:related}
In addition to the works already mentioned in the Introduction~\cite{CONGA,Drill,PLB,Flowbender,In-network,MP-RDMA,On-impact-of-PS,let-it-flow}, we briefly review a few other works that are related to our work on \sys.

\scat{Flowlets in RDMA}
As discussed earlier, RDMA traffic lacks sufficient inter-packet gaps to identify flowlets. This observation is confirmed in several works using real measurements~\cite{MP-RDMA,In-network}. So, one idea then is to artificially insert such gaps in the packet stream in order to create flowlets for load balancing, as recently proposed in $\text{HF}^2\text{T}$~\cite{HFT}. However, adding flowlet delays, specially when the network is not congested, could unnecessarily prolong FCTs and still relies on the existence of a flowlet load balancing, such as CONGA~\cite{CONGA}, in the network. 


\scat{Multipath RDMA}
MP-RDMA~\cite{MP-RDMA} includes several sophisticated mechanisms to break a flow to subflows that are then distributed over multiple paths with respect to the congestion level on those paths. However, it requires complex host-side logic to manage multiple paths and synchronize packet transmissions across them. NVIDIA collective communication library (NCCL)~\cite{nccl_website}, which is widely used for distributed training in large AI clusters, internally uses multiple QPs to send each message~\cite{meta-at-scale}, but employs random spraying similar to RPS, albeit at a chunk instead of packet granularity. Thus, in asymmetric networks, it faces similar issues as those faced by RPS.



\scat{Path Probing}
Both host and switch-based path probing have been extensively considered for various network management tasks. For example, both CONGA~\cite{CONGA} and ConWeave~\cite{In-network} implement passive and active path probing on network switches, respectively, for load balancing. On the other hand, Clove~\cite{Clove}, MP-RDMA~\cite{MP-RDMA}, and Hermes~\cite{Hermes} implement path probing on the host. For instance, Clove pre-computes a set of disjoint paths before transmission (similar to our testbed implementation) and periodically probes them, while Hermes leverages the power-of-two-choices technique to probe alternative paths with minimal overhead.


\scat{RTT Measurement}
The hardware timestamping capability of modern NICs~\cite{nvidia_connectx5, nvidia_connectx6, intelE810} has enabled precise measurement of network RTT. Timely~\cite{Timely} demonstrates that, by leveraging this capability, RTT can be measured with high precision. Swift~\cite{Swift} builds on this by using NIC-based timestamping to eliminate the effects of local NIC delays, resulting in more accurate RTT measurements. NVIDIA’s ZTR-RTT~\cite{ZTR-RTT} congestion control also relies on hardware timestamping to accurately measure RTT and dynamically adjust transmission rates.

\section{Conclusion}
\label{s:conc}

This paper presents \sys, a host-based load balancing technique for RDMA traffic in AI clusters. \sys\ operates at RTT granularity, meaning it can switch flow paths every RTT. However, to avoid unnecessarily rerouting flows, it switches paths only when the current path of a flow is congested and a less congested alternative path can be found via a proactive light-weight probing mechanism. 
Using both simulations and hardware testbed implementation, we evaluated \sys\ under traditional datacenter as well as ML training workloads. Our evaluation revealed that \sys\ outperforms state-of-the-art host-based load balancing techniques, while achieving performance close to that of switch-based techniques.

A major challenge in our work was the black-box nature of commercial RDMA NICs. To overcome this, we implemented \sys's control logic in user space. A potential direction for future research is to extend our user space implementation by incorporating \sys\ in a collective communication library, as inter-GPU communication in distributed ML training is entirely based on collective operations regardless of the model architecture or training framework. 

\bibliographystyle{ACM-Reference-Format}
\bibliography{bib/main}


\begin{thebibliography}{46}


\ifx \showCODEN    \undefined \def \showCODEN     #1{\unskip}     \fi
\ifx \showDOI      \undefined \def \showDOI       #1{#1}\fi
\ifx \showISBNx    \undefined \def \showISBNx     #1{\unskip}     \fi
\ifx \showISBNxiii \undefined \def \showISBNxiii  #1{\unskip}     \fi
\ifx \showISSN     \undefined \def \showISSN      #1{\unskip}     \fi
\ifx \showLCCN     \undefined \def \showLCCN      #1{\unskip}     \fi
\ifx \shownote     \undefined \def \shownote      #1{#1}          \fi
\ifx \showarticletitle \undefined \def \showarticletitle #1{#1}   \fi
\ifx \showURL      \undefined \def \showURL       {\relax}        \fi
\providecommand\bibfield[2]{#2}
\providecommand\bibinfo[2]{#2}
\providecommand\natexlab[1]{#1}
\providecommand\showeprint[2][]{arXiv:#2}

\bibitem[Alizadeh et~al\mbox{.}(2014)]%
        {CONGA}
\bibfield{author}{\bibinfo{person}{Mohammad Alizadeh}, \bibinfo{person}{Tom
  Edsall}, \bibinfo{person}{Sarang Dharmapurikar}, \bibinfo{person}{Ramanan
  Vaidyanathan}, \bibinfo{person}{Kevin Chu}, \bibinfo{person}{Andy Fingerhut},
  \bibinfo{person}{Vinh~The Lam}, \bibinfo{person}{Francis Matus},
  \bibinfo{person}{Rong Pan}, \bibinfo{person}{Navindra Yadav}, {and}
  \bibinfo{person}{George Varghese}.} \bibinfo{year}{2014}\natexlab{}.
\newblock \showarticletitle{CONGA: distributed congestion-aware load balancing
  for datacenters}. In \bibinfo{booktitle}{\emph{Proceedings of the 2014 ACM
  Conference on SIGCOMM}} (Chicago, Illinois, USA)
  \emph{(\bibinfo{series}{SIGCOMM '14})}. \bibinfo{publisher}{Association for
  Computing Machinery}, \bibinfo{address}{New York, NY, USA},
  \bibinfo{pages}{503–514}.
\newblock
\showISBNx{9781450328364}


\bibitem[Association(2007)]%
        {infiniband}
\bibfield{author}{\bibinfo{person}{InfiniBand~Trade Association}.}
  \bibinfo{year}{2007}\natexlab{}.
\newblock \bibinfo{title}{Infiniband Architecture Specifications}.
\newblock
  \bibinfo{howpublished}{\url{https://www.infinibandta.org/ibta-specification/}}.
\newblock


\bibitem[Cao et~al\mbox{.}(2024)]%
        {crux-cao}
\bibfield{author}{\bibinfo{person}{Jiamin Cao}, \bibinfo{person}{Yu Guan},
  \bibinfo{person}{Kun Qian}, \bibinfo{person}{Jiaqi Gao},
  \bibinfo{person}{Wencong Xiao}, \bibinfo{person}{Jianbo Dong},
  \bibinfo{person}{Binzhang Fu}, \bibinfo{person}{Dennis Cai}, {and}
  \bibinfo{person}{Ennan Zhai}.} \bibinfo{year}{2024}\natexlab{}.
\newblock \showarticletitle{Crux: GPU-Efficient Communication Scheduling for
  Deep Learning Training}. In \bibinfo{booktitle}{\emph{Proceedings of the
  Conference of the ACM Special Interest Group on Data Communication}} (Sydney,
  NSW, Australia) \emph{(\bibinfo{series}{ACM SIGCOMM '24})}.
  \bibinfo{publisher}{Association for Computing Machinery},
  \bibinfo{address}{New York, NY, USA}, \bibinfo{pages}{1–15}.
\newblock
\showISBNx{9798400706141}


\bibitem[Chen et~al\mbox{.}(2024b)]%
        {HFT}
\bibfield{author}{\bibinfo{person}{Chuhao Chen}, \bibinfo{person}{Jiarui Ye},
  \bibinfo{person}{Yongbo Gao}, \bibinfo{person}{Sen Liu}, {and}
  \bibinfo{person}{Yang Xu}.} \bibinfo{year}{2024}\natexlab{b}.
\newblock \showarticletitle{HF\^{}2T: Host-Based Flowlet Fine-Tuning for RDMA
  Load Balancing}. In \bibinfo{booktitle}{\emph{Proceedings of the 8th
  Asia-Pacific Workshop on Networking}} (Sydney, Australia)
  \emph{(\bibinfo{series}{APNet '24})}. \bibinfo{publisher}{Association for
  Computing Machinery}, \bibinfo{address}{New York, NY, USA},
  \bibinfo{pages}{9–15}.
\newblock
\showISBNx{9798400717581}


\bibitem[Chen et~al\mbox{.}(2024a)]%
        {Packet-trimming}
\bibfield{author}{\bibinfo{person}{Xiaoqi Chen}, \bibinfo{person}{Shay
  Vargaftik}, {and} \bibinfo{person}{Ran~Ben Basat}.}
  \bibinfo{year}{2024}\natexlab{a}.
\newblock \showarticletitle{When ML Training Cuts Through Congestion:
  Just-in-Time Gradient Compression via Packet Trimming}. In
  \bibinfo{booktitle}{\emph{Proceedings of the 23rd ACM Workshop on Hot Topics
  in Networks}} (Irvine, CA, USA) \emph{(\bibinfo{series}{HotNets '24})}.
  \bibinfo{publisher}{Association for Computing Machinery},
  \bibinfo{address}{New York, NY, USA}, \bibinfo{pages}{177–185}.
\newblock
\showISBNx{9798400712722}


\bibitem[Corporation(2025)]%
        {nccl_website}
\bibfield{author}{\bibinfo{person}{NVIDIA Corporation}.}
  \bibinfo{year}{2025}\natexlab{}.
\newblock \bibinfo{title}{NVIDIA Collective Communications Library (NCCL)}.
\newblock \bibinfo{howpublished}{https://developer.nvidia.com/nccl}.
\newblock
\newblock
\shownote{Accessed: 2025-05-29}.


\bibitem[Dixit et~al\mbox{.}(2013)]%
        {On-impact-of-PS}
\bibfield{author}{\bibinfo{person}{Advait Dixit}, \bibinfo{person}{Pawan
  Prakash}, \bibinfo{person}{Y.~Charlie Hu}, {and} \bibinfo{person}{Ramana~Rao
  Kompella}.} \bibinfo{year}{2013}\natexlab{}.
\newblock \showarticletitle{On the impact of packet spraying in data center
  networks}. In \bibinfo{booktitle}{\emph{2013 Proceedings IEEE INFOCOM}}.
  \bibinfo{pages}{2130--2138}.
\newblock


\bibitem[Gangidi et~al\mbox{.}(2024)]%
        {meta-at-scale}
\bibfield{author}{\bibinfo{person}{Adithya Gangidi}, \bibinfo{person}{Rui
  Miao}, \bibinfo{person}{Shengbao Zheng}, \bibinfo{person}{Sai~Jayesh Bondu},
  \bibinfo{person}{Guilherme Goes}, \bibinfo{person}{Hany Morsy},
  \bibinfo{person}{Rohit Puri}, \bibinfo{person}{Mohammad Riftadi},
  \bibinfo{person}{Ashmitha~Jeevaraj Shetty}, \bibinfo{person}{Jingyi Yang},
  \bibinfo{person}{Shuqiang Zhang}, \bibinfo{person}{Mikel~Jimenez Fernandez},
  \bibinfo{person}{Shashidhar Gandham}, {and} \bibinfo{person}{Hongyi Zeng}.}
  \bibinfo{year}{2024}\natexlab{}.
\newblock \showarticletitle{RDMA over Ethernet for Distributed Training at Meta
  Scale}. In \bibinfo{booktitle}{\emph{Proceedings of the ACM SIGCOMM 2024
  Conference}} (Sydney, NSW, Australia) \emph{(\bibinfo{series}{ACM SIGCOMM
  '24})}. \bibinfo{publisher}{Association for Computing Machinery},
  \bibinfo{address}{New York, NY, USA}, \bibinfo{pages}{57–70}.
\newblock
\showISBNx{9798400706141}


\bibitem[Ghorbani et~al\mbox{.}(2017)]%
        {Drill}
\bibfield{author}{\bibinfo{person}{Soudeh Ghorbani}, \bibinfo{person}{Zibin
  Yang}, \bibinfo{person}{P.~Brighten Godfrey}, \bibinfo{person}{Yashar
  Ganjali}, {and} \bibinfo{person}{Amin Firoozshahian}.}
  \bibinfo{year}{2017}\natexlab{}.
\newblock \showarticletitle{DRILL: Micro Load Balancing for Low-latency Data
  Center Networks}. In \bibinfo{booktitle}{\emph{Proceedings of the Conference
  of the ACM Special Interest Group on Data Communication}} (Los Angeles, CA,
  USA) \emph{(\bibinfo{series}{SIGCOMM '17})}. \bibinfo{publisher}{Association
  for Computing Machinery}, \bibinfo{address}{New York, NY, USA},
  \bibinfo{pages}{225–238}.
\newblock
\showISBNx{9781450346535}


\bibitem[Gong et~al\mbox{.}(2018)]%
        {HeavyKeeper}
\bibfield{author}{\bibinfo{person}{Junzhi Gong}, \bibinfo{person}{Tong Yang},
  \bibinfo{person}{Haowei Zhang}, \bibinfo{person}{Hao Li},
  \bibinfo{person}{Steve Uhlig}, \bibinfo{person}{Shigang Chen},
  \bibinfo{person}{Lorna Uden}, {and} \bibinfo{person}{Xiaoming Li}.}
  \bibinfo{year}{2018}\natexlab{}.
\newblock \showarticletitle{{HeavyKeeper}: An Accurate Algorithm for Finding
  Top-k Elephant Flows}. In \bibinfo{booktitle}{\emph{2018 USENIX Annual
  Technical Conference (USENIX ATC 18)}}. \bibinfo{publisher}{USENIX
  Association}, \bibinfo{address}{Boston, MA}, \bibinfo{pages}{909--921}.
\newblock
\showISBNx{978-1-939133-01-4}


\bibitem[Greenberg et~al\mbox{.}(2009)]%
        {topo-green}
\bibfield{author}{\bibinfo{person}{Albert Greenberg}, \bibinfo{person}{James~R.
  Hamilton}, \bibinfo{person}{Navendu Jain}, \bibinfo{person}{Srikanth
  Kandula}, \bibinfo{person}{Changhoon Kim}, \bibinfo{person}{Parantap Lahiri},
  \bibinfo{person}{David~A. Maltz}, \bibinfo{person}{Parveen Patel}, {and}
  \bibinfo{person}{Sudipta Sengupta}.} \bibinfo{year}{2009}\natexlab{}.
\newblock \showarticletitle{VL2: a scalable and flexible data center network}.
  In \bibinfo{booktitle}{\emph{Proceedings of the ACM SIGCOMM 2009 Conference
  on Data Communication}} (Barcelona, Spain) \emph{(\bibinfo{series}{SIGCOMM
  '09})}. \bibinfo{publisher}{Association for Computing Machinery},
  \bibinfo{address}{New York, NY, USA}, \bibinfo{pages}{51–62}.
\newblock
\showISBNx{9781605585949}


\bibitem[Hopps(2000)]%
        {ecmp}
\bibfield{author}{\bibinfo{person}{Christian Hopps}.}
  \bibinfo{year}{2000}\natexlab{}.
\newblock \bibinfo{title}{Analysis of an Equal-Cost Multi-Path Algorithm}.
\newblock \bibinfo{howpublished}{RFC 2992}.
\newblock


\bibitem[IEEE(2008)]%
        {pfc}
\bibfield{author}{\bibinfo{person}{IEEE}.} \bibinfo{year}{2008}\natexlab{}.
\newblock \bibinfo{title}{{802.1Qbb} – Priority-based Flow Control}.
\newblock \bibinfo{howpublished}{\url{https://1.ieee802.org/dcb/802-1qbb/}}.
\newblock


\bibitem[{Intel Corporation}(2025)]%
        {intelE810}
\bibfield{author}{\bibinfo{person}{{Intel Corporation}}.}
  \bibinfo{year}{2025}\natexlab{}.
\newblock \bibinfo{title}{Intel® Ethernet Network Adapter E810-2CQDA2
  Specifications}.
\newblock
  \bibinfo{howpublished}{\url{https://www.intel.com/content/www/us/en/products/sku/210969/intel-ethernet-network-adapter-e8102cqda2/specifications.html}}.
\newblock
\newblock
\shownote{Accessed: 2025-06-02}.


\bibitem[Kabbani et~al\mbox{.}(2014)]%
        {Flowbender}
\bibfield{author}{\bibinfo{person}{Abdul Kabbani}, \bibinfo{person}{Balajee
  Vamanan}, \bibinfo{person}{Jahangir Hasan}, {and} \bibinfo{person}{Fabien
  Duchene}.} \bibinfo{year}{2014}\natexlab{}.
\newblock \showarticletitle{FlowBender: Flow-level Adaptive Routing for
  Improved Latency and Throughput in Datacenter Networks}. In
  \bibinfo{booktitle}{\emph{Proceedings of the 10th ACM International on
  Conference on Emerging Networking Experiments and Technologies}} (Sydney,
  Australia) \emph{(\bibinfo{series}{CoNEXT '14})}.
  \bibinfo{publisher}{Association for Computing Machinery},
  \bibinfo{address}{New York, NY, USA}, \bibinfo{pages}{149–160}.
\newblock
\showISBNx{9781450332798}


\bibitem[Kalia et~al\mbox{.}(2019)]%
        {rpc-kalia}
\bibfield{author}{\bibinfo{person}{Anuj Kalia}, \bibinfo{person}{Michael
  Kaminsky}, {and} \bibinfo{person}{David Andersen}.}
  \bibinfo{year}{2019}\natexlab{}.
\newblock \showarticletitle{Datacenter {RPCs} can be General and Fast}. In
  \bibinfo{booktitle}{\emph{16th USENIX Symposium on Networked Systems Design
  and Implementation (NSDI 19)}}. \bibinfo{publisher}{USENIX Association},
  \bibinfo{address}{Boston, MA}, \bibinfo{pages}{1--16}.
\newblock
\showISBNx{978-1-931971-49-2}


\bibitem[Katta et~al\mbox{.}(2017)]%
        {Clove}
\bibfield{author}{\bibinfo{person}{Naga Katta}, \bibinfo{person}{Aditi Ghag},
  \bibinfo{person}{Mukesh Hira}, \bibinfo{person}{Isaac Keslassy},
  \bibinfo{person}{Aran Bergman}, \bibinfo{person}{Changhoon Kim}, {and}
  \bibinfo{person}{Jennifer Rexford}.} \bibinfo{year}{2017}\natexlab{}.
\newblock \showarticletitle{Clove: Congestion-Aware Load Balancing at the
  Virtual Edge}. In \bibinfo{booktitle}{\emph{Proceedings of the 13th
  International Conference on Emerging Networking EXperiments and
  Technologies}} (Incheon, Republic of Korea) \emph{(\bibinfo{series}{CoNEXT
  '17})}. \bibinfo{publisher}{Association for Computing Machinery},
  \bibinfo{address}{New York, NY, USA}, \bibinfo{pages}{323–335}.
\newblock
\showISBNx{9781450354226}
\urldef\tempurl%
\url{https://doi.org/10.1145/3143361.3143401}
\showDOI{\tempurl}


\bibitem[Kumar et~al\mbox{.}(2020)]%
        {Swift}
\bibfield{author}{\bibinfo{person}{Gautam Kumar}, \bibinfo{person}{Nandita
  Dukkipati}, \bibinfo{person}{Keon Jang}, \bibinfo{person}{Hassan M.~G.
  Wassel}, \bibinfo{person}{Xian Wu}, \bibinfo{person}{Behnam Montazeri},
  \bibinfo{person}{Yaogong Wang}, \bibinfo{person}{Kevin Springborn},
  \bibinfo{person}{Christopher Alfeld}, \bibinfo{person}{Michael Ryan},
  \bibinfo{person}{David Wetherall}, {and} \bibinfo{person}{Amin Vahdat}.}
  \bibinfo{year}{2020}\natexlab{}.
\newblock \showarticletitle{Swift: Delay is Simple and Effective for Congestion
  Control in the Datacenter}. In \bibinfo{booktitle}{\emph{Proceedings of the
  Annual Conference of the ACM Special Interest Group on Data Communication on
  the Applications, Technologies, Architectures, and Protocols for Computer
  Communication}} (Virtual Event, USA) \emph{(\bibinfo{series}{SIGCOMM '20})}.
  \bibinfo{publisher}{Association for Computing Machinery},
  \bibinfo{address}{New York, NY, USA}, \bibinfo{pages}{514–528}.
\newblock
\showISBNx{9781450379557}


\bibitem[Le et~al\mbox{.}(2024)]%
        {Strack}
\bibfield{author}{\bibinfo{person}{Yanfang Le}, \bibinfo{person}{Rong Pan},
  \bibinfo{person}{Peter Newman}, \bibinfo{person}{Jeremias Blendin},
  \bibinfo{person}{Abdul Kabbani}, \bibinfo{person}{Vipin Jain},
  \bibinfo{person}{Raghava Sivaramu}, {and} \bibinfo{person}{Francis Matus}.}
  \bibinfo{year}{2024}\natexlab{}.
\newblock \bibinfo{title}{STrack: A Reliable Multipath Transport for AI/ML
  Clusters}.
\newblock
\newblock
\showeprint[arxiv]{2407.15266}


\bibitem[Li et~al\mbox{.}(2020)]%
        {ddp}
\bibfield{author}{\bibinfo{person}{Shen Li}, \bibinfo{person}{Yanli Zhao},
  \bibinfo{person}{Rohan Varma}, \bibinfo{person}{Omkar Salpekar},
  \bibinfo{person}{Pieter Noordhuis}, \bibinfo{person}{Teng Li},
  \bibinfo{person}{Adam Paszke}, \bibinfo{person}{Jeff Smith},
  \bibinfo{person}{Brian Vaughan}, \bibinfo{person}{Pritam Damania}, {and}
  \bibinfo{person}{Soumith Chintala}.} \bibinfo{year}{2020}\natexlab{}.
\newblock \bibinfo{title}{PyTorch Distributed: Experiences on Accelerating Data
  Parallel Training}.
\newblock
\newblock
\showeprint[arxiv]{2006.15704}~[cs.DC]
\urldef\tempurl%
\url{https://arxiv.org/abs/2006.15704}
\showURL{%
\tempurl}


\bibitem[Li et~al\mbox{.}(2019)]%
        {HPCC}
\bibfield{author}{\bibinfo{person}{Yuliang Li}, \bibinfo{person}{Rui Miao},
  \bibinfo{person}{Hongqiang~Harry Liu}, \bibinfo{person}{Yan Zhuang},
  \bibinfo{person}{Fei Feng}, \bibinfo{person}{Lingbo Tang},
  \bibinfo{person}{Zheng Cao}, \bibinfo{person}{Ming Zhang},
  \bibinfo{person}{Frank Kelly}, \bibinfo{person}{Mohammad Alizadeh}, {and}
  \bibinfo{person}{Minlan Yu}.} \bibinfo{year}{2019}\natexlab{}.
\newblock \showarticletitle{HPCC: high precision congestion control}. In
  \bibinfo{booktitle}{\emph{Proceedings of the ACM Special Interest Group on
  Data Communication}} (Beijing, China) \emph{(\bibinfo{series}{SIGCOMM '19})}.
  \bibinfo{publisher}{Association for Computing Machinery},
  \bibinfo{address}{New York, NY, USA}, \bibinfo{pages}{44–58}.
\newblock
\showISBNx{9781450359566}


\bibitem[{Linux RDMA}(2025)]%
        {rdma-core}
\bibfield{author}{\bibinfo{person}{{Linux RDMA}}.}
  \bibinfo{year}{2025}\natexlab{}.
\newblock \bibinfo{title}{RDMA core userspace libraries and daemons}.
\newblock
  \bibinfo{howpublished}{\url{https://github.com/linux-rdma/rdma-core}}.
\newblock
\newblock
\shownote{Accessed: 2025-05-16}.


\bibitem[Long et~al\mbox{.}(2024)]%
        {LSCC}
\bibfield{author}{\bibinfo{person}{Minfei Long}, \bibinfo{person}{Jiangping
  Han}, \bibinfo{person}{Wentao Wang}, \bibinfo{person}{Jiayu Yang}, {and}
  \bibinfo{person}{Kaiping Xue}.} \bibinfo{year}{2024}\natexlab{}.
\newblock \showarticletitle{LSCC: Link-Segmented Congestion Control for RDMA in
  Cross-Datacenter Networks}. In \bibinfo{booktitle}{\emph{2024 IEEE/ACM 32nd
  International Symposium on Quality of Service (IWQoS)}}.
  \bibinfo{pages}{1--10}.
\newblock
\urldef\tempurl%
\url{https://doi.org/10.1109/IWQoS61813.2024.10682909}
\showDOI{\tempurl}


\bibitem[Lu et~al\mbox{.}(2018)]%
        {MP-RDMA}
\bibfield{author}{\bibinfo{person}{Yuanwei Lu}, \bibinfo{person}{Guo Chen},
  \bibinfo{person}{Bojie Li}, \bibinfo{person}{Kun Tan},
  \bibinfo{person}{Yongqiang Xiong}, \bibinfo{person}{Peng Cheng},
  \bibinfo{person}{Jiansong Zhang}, \bibinfo{person}{Enhong Chen}, {and}
  \bibinfo{person}{Thomas Moscibroda}.} \bibinfo{year}{2018}\natexlab{}.
\newblock \showarticletitle{{Multi-Path} Transport for {RDMA} in Datacenters}.
  In \bibinfo{booktitle}{\emph{15th USENIX Symposium on Networked Systems
  Design and Implementation (NSDI 18)}}. \bibinfo{publisher}{USENIX
  Association}, \bibinfo{address}{Renton, WA}, \bibinfo{pages}{357--371}.
\newblock
\showISBNx{978-1-939133-01-4}


\bibitem[Mittal et~al\mbox{.}(2015)]%
        {Timely}
\bibfield{author}{\bibinfo{person}{Radhika Mittal}, \bibinfo{person}{Vinh~The
  Lam}, \bibinfo{person}{Nandita Dukkipati}, \bibinfo{person}{Emily Blem},
  \bibinfo{person}{Hassan Wassel}, \bibinfo{person}{Monia Ghobadi},
  \bibinfo{person}{Amin Vahdat}, \bibinfo{person}{Yaogong Wang},
  \bibinfo{person}{David Wetherall}, {and} \bibinfo{person}{David Zats}.}
  \bibinfo{year}{2015}\natexlab{}.
\newblock \showarticletitle{TIMELY: RTT-based Congestion Control for the
  Datacenter}. In \bibinfo{booktitle}{\emph{Proceedings of the 2015 ACM
  Conference on Special Interest Group on Data Communication}} (London, United
  Kingdom) \emph{(\bibinfo{series}{SIGCOMM '15})}.
  \bibinfo{publisher}{Association for Computing Machinery},
  \bibinfo{address}{New York, NY, USA}, \bibinfo{pages}{537–550}.
\newblock
\showISBNx{9781450335423}


\bibitem[Mittal et~al\mbox{.}(2018)]%
        {RDMA-IRN}
\bibfield{author}{\bibinfo{person}{Radhika Mittal}, \bibinfo{person}{Alexander
  Shpiner}, \bibinfo{person}{Aurojit Panda}, \bibinfo{person}{Eitan Zahavi},
  \bibinfo{person}{Arvind Krishnamurthy}, \bibinfo{person}{Sylvia Ratnasamy},
  {and} \bibinfo{person}{Scott Shenker}.} \bibinfo{year}{2018}\natexlab{}.
\newblock \showarticletitle{Revisiting network support for RDMA}. In
  \bibinfo{booktitle}{\emph{Proceedings of the 2018 Conference of the ACM
  Special Interest Group on Data Communication}} (Budapest, Hungary)
  \emph{(\bibinfo{series}{SIGCOMM '18})}. \bibinfo{publisher}{Association for
  Computing Machinery}, \bibinfo{address}{New York, NY, USA},
  \bibinfo{pages}{313–326}.
\newblock
\showISBNx{9781450355674}


\bibitem[{ns-3 Project}(2025)]%
        {ns3_website}
\bibfield{author}{\bibinfo{person}{{ns-3 Project}}.}
  \bibinfo{year}{2025}\natexlab{}.
\newblock \bibinfo{title}{{ns-3}: A Discrete-Event Network Simulator}.
\newblock \bibinfo{howpublished}{\url{https://www.nsnam.org/}}.
\newblock
\newblock
\shownote{Accessed: 2025-05-02}.


\bibitem[{NVIDIA}(2024)]%
        {ZTR-RTT}
\bibfield{author}{\bibinfo{person}{{NVIDIA}}.} \bibinfo{year}{2024}\natexlab{}.
\newblock \bibinfo{title}{ZTR-RTT Congestion Control Algorithm Overview v1.0}.
\newblock
\newblock
\urldef\tempurl%
\url{https://docs.nvidia.com/networking/display/ztrrttcongestioncontrolalgorithmoverviewv10}
\showURL{%
\tempurl}
\newblock
\shownote{Accessed: 2025-05-16}.


\bibitem[NVIDIA(nda)]%
        {nvidia_connectx5}
\bibfield{author}{\bibinfo{person}{NVIDIA}.}
  \bibinfo{year}{[n.d.]}\natexlab{a}.
\newblock \bibinfo{title}{ConnectX-5 SmartNIC Adapter}.
\newblock
  \bibinfo{howpublished}{\url{https://www.nvidia.com/en-sg/networking/ethernet/connectx-5/}}.
\newblock
\newblock
\shownote{Accessed: 2025-05-28}.


\bibitem[NVIDIA(ndb)]%
        {nvidia_connectx6}
\bibfield{author}{\bibinfo{person}{NVIDIA}.}
  \bibinfo{year}{[n.d.]}\natexlab{b}.
\newblock \bibinfo{title}{ConnectX-6 SmartNIC Adapter}.
\newblock
  \bibinfo{howpublished}{\url{https://www.nvidia.com/en-sg/networking/ethernet/connectx-6/}}.
\newblock
\newblock
\shownote{Accessed: 2025-05-28}.


\bibitem[Project(2025a)]%
        {dpdk_website}
\bibfield{author}{\bibinfo{person}{DPDK Project}.}
  \bibinfo{year}{2025}\natexlab{a}.
\newblock \bibinfo{title}{Data Plane Development Kit (DPDK)}.
\newblock \bibinfo{howpublished}{\url{https://www.dpdk.org/}}.
\newblock
\newblock
\shownote{Accessed: 2025-05-02}.


\bibitem[Project(2025b)]%
        {sonic2025}
\bibfield{author}{\bibinfo{person}{{SONiC} Project}.}
  \bibinfo{year}{2025}\natexlab{b}.
\newblock \bibinfo{title}{Software for Open Networking in the Cloud}.
\newblock \bibinfo{howpublished}{\url{https://sonicfoundation.dev/}}.
\newblock
\newblock
\shownote{Accessed: 2025-06-28}.


\bibitem[Qian et~al\mbox{.}(2024)]%
        {HPN}
\bibfield{author}{\bibinfo{person}{Kun Qian}, \bibinfo{person}{Yongqing Xi},
  \bibinfo{person}{Jiamin Cao}, \bibinfo{person}{Jiaqi Gao},
  \bibinfo{person}{Yichi Xu}, \bibinfo{person}{Yu Guan},
  \bibinfo{person}{Binzhang Fu}, \bibinfo{person}{Xuemei Shi},
  \bibinfo{person}{Fangbo Zhu}, \bibinfo{person}{Rui Miao},
  \bibinfo{person}{Chao Wang}, \bibinfo{person}{Peng Wang},
  \bibinfo{person}{Pengcheng Zhang}, \bibinfo{person}{Xianlong Zeng},
  \bibinfo{person}{Eddie Ruan}, \bibinfo{person}{Zhiping Yao},
  \bibinfo{person}{Ennan Zhai}, {and} \bibinfo{person}{Dennis Cai}.}
  \bibinfo{year}{2024}\natexlab{}.
\newblock \showarticletitle{Alibaba HPN: A Data Center Network for Large
  Language Model Training}. In \bibinfo{booktitle}{\emph{Proceedings of the ACM
  SIGCOMM 2024 Conference}} (Sydney, NSW, Australia)
  \emph{(\bibinfo{series}{ACM SIGCOMM '24})}. \bibinfo{publisher}{Association
  for Computing Machinery}, \bibinfo{address}{New York, NY, USA},
  \bibinfo{pages}{691–706}.
\newblock
\showISBNx{9798400706141}


\bibitem[Qureshi et~al\mbox{.}(2022)]%
        {PLB}
\bibfield{author}{\bibinfo{person}{Mubashir~Adnan Qureshi},
  \bibinfo{person}{Yuchung Cheng}, \bibinfo{person}{Qianwen Yin},
  \bibinfo{person}{Qiaobin Fu}, \bibinfo{person}{Gautam Kumar},
  \bibinfo{person}{Masoud Moshref}, \bibinfo{person}{Junhua Yan},
  \bibinfo{person}{Van Jacobson}, \bibinfo{person}{David Wetherall}, {and}
  \bibinfo{person}{Abdul Kabbani}.} \bibinfo{year}{2022}\natexlab{}.
\newblock \showarticletitle{PLB: congestion signals are simple and effective
  for network load balancing}. In \bibinfo{booktitle}{\emph{Proceedings of the
  Conference of the ACM Special Interest Group on Data Communication}}
  (Amsterdam, Netherlands) \emph{(\bibinfo{series}{ACM SIGCOMM '22})}.
  \bibinfo{publisher}{Association for Computing Machinery},
  \bibinfo{address}{New York, NY, USA}, \bibinfo{pages}{207–218}.
\newblock
\showISBNx{9781450394208}


\bibitem[Ravi et~al\mbox{.}(2024)]%
        {csig}
\bibfield{author}{\bibinfo{person}{Abhiram Ravi}, \bibinfo{person}{Nandita
  Dukkipati}, \bibinfo{person}{Naoshad Mehta}, {and} \bibinfo{person}{Jai
  Kumar}.} \bibinfo{year}{2024}\natexlab{}.
\newblock \bibinfo{booktitle}{\emph{{Congestion Signaling (CSIG)}}}.
\newblock \bibinfo{type}{Internet-Draft} draft-ravi-ippm-csig-01.
  \bibinfo{institution}{Internet Engineering Task Force}.
\newblock
\urldef\tempurl%
\url{https://datatracker.ietf.org/doc/draft-ravi-ippm-csig/01/}
\showURL{%
\tempurl}
\newblock
\shownote{Work in Progress}.


\bibitem[Roy et~al\mbox{.}(2015)]%
        {Meta-hadoop}
\bibfield{author}{\bibinfo{person}{Arjun Roy}, \bibinfo{person}{Hongyi Zeng},
  \bibinfo{person}{Jasmeet Bagga}, \bibinfo{person}{George Porter}, {and}
  \bibinfo{person}{Alex~C. Snoeren}.} \bibinfo{year}{2015}\natexlab{}.
\newblock \showarticletitle{Inside the Social Network's (Datacenter) Network}.
\newblock \bibinfo{journal}{\emph{SIGCOMM Comput. Commun. Rev.}}
  \bibinfo{volume}{45}, \bibinfo{number}{4} (\bibinfo{date}{Aug.}
  \bibinfo{year}{2015}), \bibinfo{pages}{123–137}.
\newblock
\showISSN{0146-4833}


\bibitem[Song et~al\mbox{.}(2023)]%
        {In-network}
\bibfield{author}{\bibinfo{person}{Cha~Hwan Song}, \bibinfo{person}{Xin~Zhe
  Khooi}, \bibinfo{person}{Raj Joshi}, \bibinfo{person}{Inho Choi},
  \bibinfo{person}{Jialin Li}, {and} \bibinfo{person}{Mun~Choon Chan}.}
  \bibinfo{year}{2023}\natexlab{}.
\newblock \showarticletitle{Network Load Balancing with In-network Reordering
  Support for RDMA}. In \bibinfo{booktitle}{\emph{Proceedings of the Conference
  of the ACM Special Interest Group on Data Communication}} (New York, NY, USA)
  \emph{(\bibinfo{series}{ACM SIGCOMM '23})}. \bibinfo{publisher}{Association
  for Computing Machinery}, \bibinfo{address}{New York, NY, USA},
  \bibinfo{pages}{816–831}.
\newblock
\showISBNx{9798400702365}


\bibitem[Vanini et~al\mbox{.}(2017)]%
        {let-it-flow}
\bibfield{author}{\bibinfo{person}{Erico Vanini}, \bibinfo{person}{Rong Pan},
  \bibinfo{person}{Mohammad Alizadeh}, \bibinfo{person}{Parvin Taheri}, {and}
  \bibinfo{person}{Tom Edsall}.} \bibinfo{year}{2017}\natexlab{}.
\newblock \showarticletitle{Let It Flow: Resilient Asymmetric Load Balancing
  with Flowlet Switching}. In \bibinfo{booktitle}{\emph{14th USENIX Symposium
  on Networked Systems Design and Implementation (NSDI 17)}}.
  \bibinfo{publisher}{USENIX Association}, \bibinfo{address}{Boston, MA},
  \bibinfo{pages}{407--420}.
\newblock
\showISBNx{978-1-931971-37-9}


\bibitem[Wan et~al\mbox{.}(2024)]%
        {BiCC}
\bibfield{author}{\bibinfo{person}{Zirui Wan}, \bibinfo{person}{Jiao Zhang},
  \bibinfo{person}{Mingxuan Yu}, \bibinfo{person}{Junwei Liu},
  \bibinfo{person}{Jun Yao}, \bibinfo{person}{Xinghua Zhao}, {and}
  \bibinfo{person}{Tao Huang}.} \bibinfo{year}{2024}\natexlab{}.
\newblock \showarticletitle{BiCC: Bilateral Congestion Control in
  Cross-datacenter RDMA Networks}. In \bibinfo{booktitle}{\emph{IEEE INFOCOM
  2024 - IEEE Conference on Computer Communications}}.
  \bibinfo{pages}{1381--1390}.
\newblock
\urldef\tempurl%
\url{https://doi.org/10.1109/INFOCOM52122.2024.10621412}
\showDOI{\tempurl}


\bibitem[Wang et~al\mbox{.}(2023)]%
        {SRNIC}
\bibfield{author}{\bibinfo{person}{Zilong Wang}, \bibinfo{person}{Layong Luo},
  \bibinfo{person}{Qingsong Ning}, \bibinfo{person}{Chaoliang Zeng},
  \bibinfo{person}{Wenxue Li}, \bibinfo{person}{Xinchen Wan},
  \bibinfo{person}{Peng Xie}, \bibinfo{person}{Tao Feng}, \bibinfo{person}{Ke
  Cheng}, \bibinfo{person}{Xiongfei Geng}, \bibinfo{person}{Tianhao Wang},
  \bibinfo{person}{Weicheng Ling}, \bibinfo{person}{Kejia Huo},
  \bibinfo{person}{Pingbo An}, \bibinfo{person}{Kui Ji},
  \bibinfo{person}{Shideng Zhang}, \bibinfo{person}{Bin Xu},
  \bibinfo{person}{Ruiqing Feng}, \bibinfo{person}{Tao Ding},
  \bibinfo{person}{Kai Chen}, {and} \bibinfo{person}{Chuanxiong Guo}.}
  \bibinfo{year}{2023}\natexlab{}.
\newblock \showarticletitle{{SRNIC}: A Scalable Architecture for {RDMA}
  {NICs}}. In \bibinfo{booktitle}{\emph{20th USENIX Symposium on Networked
  Systems Design and Implementation (NSDI 23)}}. \bibinfo{publisher}{USENIX
  Association}, \bibinfo{address}{Boston, MA}, \bibinfo{pages}{1--14}.
\newblock
\showISBNx{978-1-939133-33-5}


\bibitem[Won et~al\mbox{.}(2023)]%
        {Astrasim2}
\bibfield{author}{\bibinfo{person}{William Won}, \bibinfo{person}{Taekyung
  Heo}, \bibinfo{person}{Saeed Rashidi}, \bibinfo{person}{Srinivas Sridharan},
  \bibinfo{person}{Sudarshan Srinivasan}, {and} \bibinfo{person}{Tushar
  Krishna}.} \bibinfo{year}{2023}\natexlab{}.
\newblock \showarticletitle{ASTRA-sim2.0: Modeling Hierarchical Networks and
  Disaggregated Systems for Large-model Training at Scale}. In
  \bibinfo{booktitle}{\emph{2023 IEEE International Symposium on Performance
  Analysis of Systems and Software (ISPASS)}}. \bibinfo{publisher}{IEEE},
  \bibinfo{address}{New York, NY, USA}, \bibinfo{pages}{283–294}.
\newblock


\bibitem[Xu et~al\mbox{.}(2022)]%
        {hash-xu}
\bibfield{author}{\bibinfo{person}{Yunhong Xu}, \bibinfo{person}{Keqiang He},
  \bibinfo{person}{Rui Wang}, \bibinfo{person}{Minlan Yu},
  \bibinfo{person}{Nick Duffield}, \bibinfo{person}{Hassan Wassel},
  \bibinfo{person}{Shidong Zhang}, \bibinfo{person}{Leon Poutievski},
  \bibinfo{person}{Junlan Zhou}, {and} \bibinfo{person}{Amin Vahdat}.}
  \bibinfo{year}{2022}\natexlab{}.
\newblock \showarticletitle{Hashing Design in Modern Networks: Challenges and
  Mitigation Techniques}. In \bibinfo{booktitle}{\emph{2022 USENIX Annual
  Technical Conference (USENIX ATC 22)}}. \bibinfo{publisher}{USENIX
  Association}, \bibinfo{address}{Carlsbad, CA}, \bibinfo{pages}{805--818}.
\newblock
\showISBNx{978-1-939133-29-45}


\bibitem[Zhang et~al\mbox{.}(2017)]%
        {Hermes}
\bibfield{author}{\bibinfo{person}{Hong Zhang}, \bibinfo{person}{Junxue Zhang},
  \bibinfo{person}{Wei Bai}, \bibinfo{person}{Kai Chen}, {and}
  \bibinfo{person}{Mosharaf Chowdhury}.} \bibinfo{year}{2017}\natexlab{}.
\newblock \showarticletitle{Resilient Datacenter Load Balancing in the Wild}.
  In \bibinfo{booktitle}{\emph{Proceedings of the Conference of the ACM Special
  Interest Group on Data Communication}} (Los Angeles, CA, USA)
  \emph{(\bibinfo{series}{SIGCOMM '17})}. \bibinfo{publisher}{Association for
  Computing Machinery}, \bibinfo{address}{New York, NY, USA},
  \bibinfo{pages}{253–266}.
\newblock
\showISBNx{9781450346535}
\urldef\tempurl%
\url{https://doi.org/10.1145/3098822.3098841}
\showDOI{\tempurl}


\bibitem[Zhang et~al\mbox{.}(2024)]%
        {PACC}
\bibfield{author}{\bibinfo{person}{Jiao Zhang}, \bibinfo{person}{Yuqing Wang},
  \bibinfo{person}{Xiaolong Zhong}, \bibinfo{person}{Mingxuan Yu},
  \bibinfo{person}{Haoyu Pan}, \bibinfo{person}{Yali Zhang},
  \bibinfo{person}{Zixuan Guan}, \bibinfo{person}{Biyao Che},
  \bibinfo{person}{Zirui Wan}, \bibinfo{person}{Tian Pan}, {and}
  \bibinfo{person}{Tao Huang}.} \bibinfo{year}{2024}\natexlab{}.
\newblock \showarticletitle{PACC: A Proactive CNP Generation Scheme for
  Datacenter Networks}.
\newblock \bibinfo{journal}{\emph{IEEE/ACM Trans. Netw.}} \bibinfo{volume}{32},
  \bibinfo{number}{3} (\bibinfo{date}{Feb.} \bibinfo{year}{2024}),
  \bibinfo{pages}{2586–2599}.
\newblock
\showISSN{1063-6692}
\urldef\tempurl%
\url{https://doi.org/10.1109/TNET.2024.3361771}
\showDOI{\tempurl}


\bibitem[Zhao et~al\mbox{.}(2023)]%
        {fsdp}
\bibfield{author}{\bibinfo{person}{Yanli Zhao}, \bibinfo{person}{Andrew Gu},
  \bibinfo{person}{Rohan Varma}, \bibinfo{person}{Liang Luo},
  \bibinfo{person}{Chien-Chin Huang}, \bibinfo{person}{Min Xu},
  \bibinfo{person}{Less Wright}, \bibinfo{person}{Hamid Shojanazeri},
  \bibinfo{person}{Myle Ott}, \bibinfo{person}{Sam Shleifer},
  \bibinfo{person}{Alban Desmaison}, \bibinfo{person}{Can Balioglu},
  \bibinfo{person}{Pritam Damania}, \bibinfo{person}{Bernard Nguyen},
  \bibinfo{person}{Geeta Chauhan}, \bibinfo{person}{Yuchen Hao},
  \bibinfo{person}{Ajit Mathews}, {and} \bibinfo{person}{Shen Li}.}
  \bibinfo{year}{2023}\natexlab{}.
\newblock \bibinfo{title}{PyTorch FSDP: Experiences on Scaling Fully Sharded
  Data Parallel}.
\newblock
\newblock
\showeprint[arxiv]{2304.11277}~[cs.DC]
\urldef\tempurl%
\url{https://arxiv.org/abs/2304.11277}
\showURL{%
\tempurl}


\bibitem[Zhu et~al\mbox{.}(2015)]%
        {DCQCN}
\bibfield{author}{\bibinfo{person}{Yibo Zhu}, \bibinfo{person}{Haggai Eran},
  \bibinfo{person}{Daniel Firestone}, \bibinfo{person}{Chuanxiong Guo},
  \bibinfo{person}{Marina Lipshteyn}, \bibinfo{person}{Yehonatan Liron},
  \bibinfo{person}{Jitendra Padhye}, \bibinfo{person}{Shachar Raindel},
  \bibinfo{person}{Mohamad~Haj Yahia}, {and} \bibinfo{person}{Ming Zhang}.}
  \bibinfo{year}{2015}\natexlab{}.
\newblock \showarticletitle{Congestion Control for Large-Scale RDMA
  Deployments}. In \bibinfo{booktitle}{\emph{Proceedings of the 2015 ACM
  Conference on Special Interest Group on Data Communication}} (London, United
  Kingdom) \emph{(\bibinfo{series}{SIGCOMM '15})}.
  \bibinfo{publisher}{Association for Computing Machinery},
  \bibinfo{address}{New York, NY, USA}, \bibinfo{pages}{523–536}.
\newblock
\showISBNx{9781450335423}


\end{thebibliography}

\newpage

\appendix 

\section{Hopper's Workflow}
\label{a:workflow}
\begin{figure*}[t]
	\centering
	
	\begin{subfigure}[t]{0.48\textwidth}
		\centering
		\includegraphics[width=\textwidth]{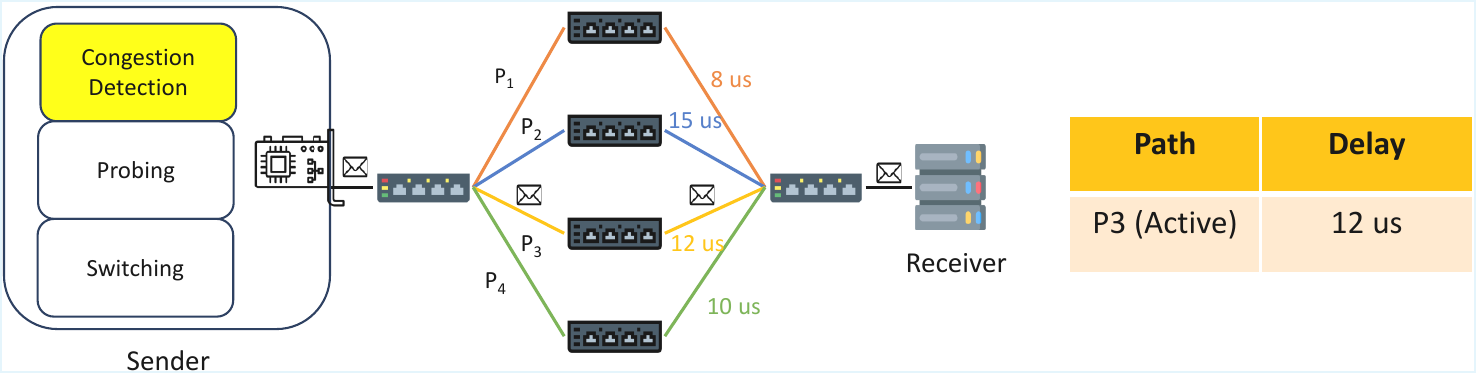}
		\caption{\scriptsize Detecting congestion.}
		\label{fig:hopper:congestion}
	\end{subfigure}
	\hfill
	\begin{subfigure}[t]{0.48\textwidth}
		\centering
		\includegraphics[width=\textwidth]{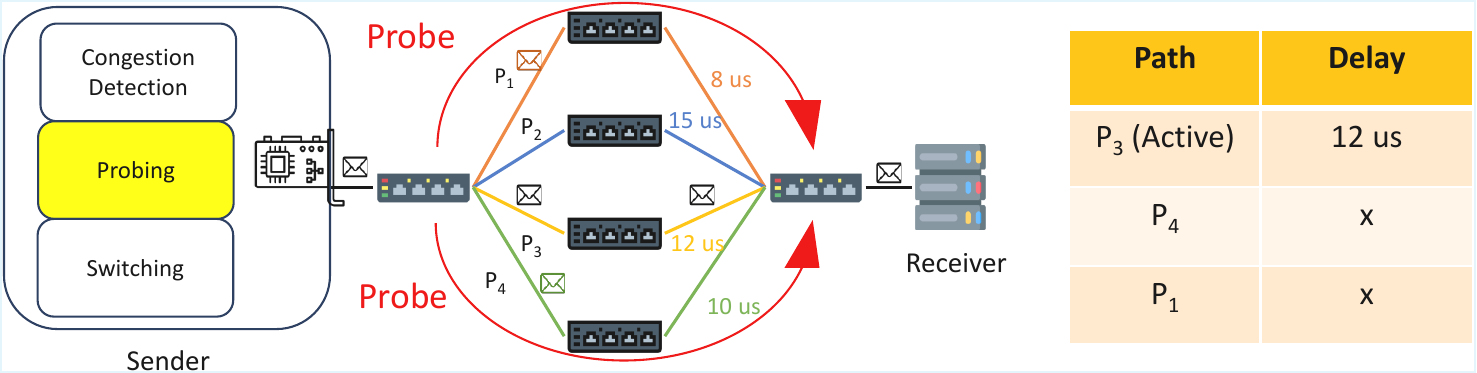}
		\caption{\scriptsize Probing alternative paths.}
		\label{fig:hopper:probing}
	\end{subfigure}
	
	\vspace{2mm} 
	
	\begin{subfigure}[t]{0.48\textwidth}
		\centering
		\includegraphics[width=\textwidth]{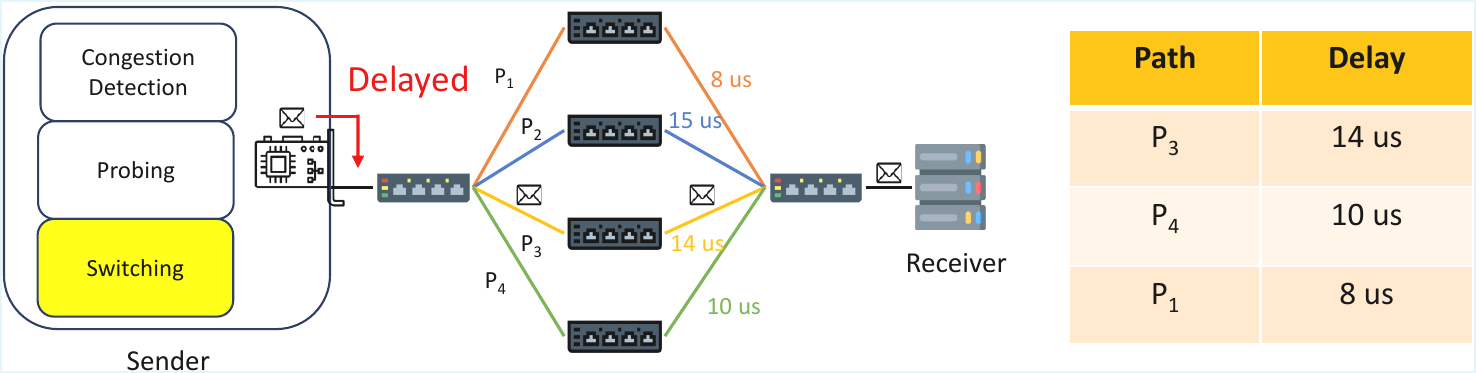}
		\caption{\scriptsize Delaying switching.}
		\label{fig:hopper:delaying}
	\end{subfigure}
	\hfill
	\begin{subfigure}[t]{0.48\textwidth}
		\centering
		\includegraphics[width=\textwidth]{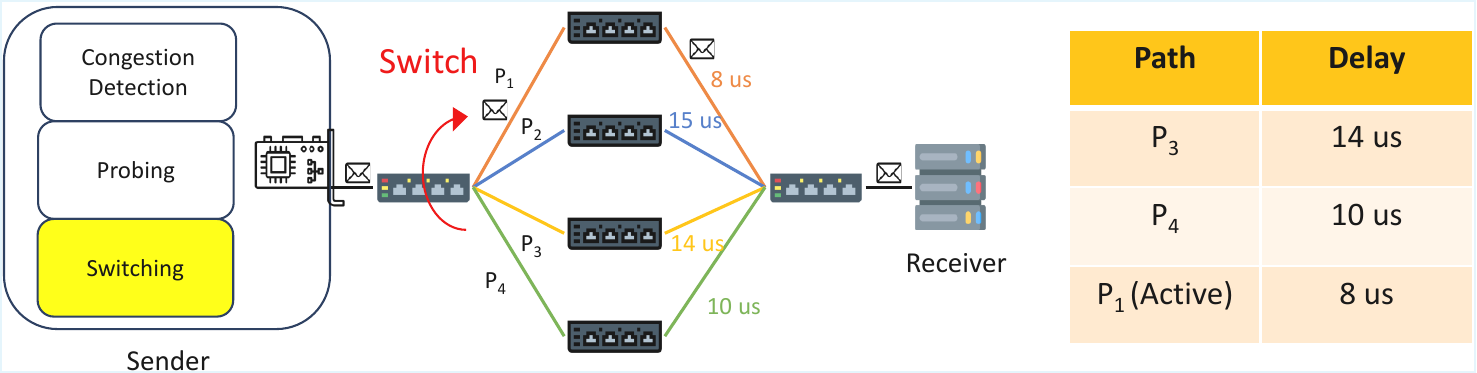}
		\caption{\scriptsize Path switching.}
		\label{fig:hopper:switching}
	\end{subfigure}
	
	\caption{Hoppers workflow.}
	\label{fig:hopper:workflow}
\end{figure*}

The workflow of \sys\ is illustrated in Figure~\ref{fig:hopper:workflow}. Below, we explain the \sys's workflow in detail.
\begin{itemize}
	\item The congestion detection module actively monitors the RTT of the current path ($\text{P}_3$) (see Fig.~\ref{fig:hopper:congestion}).
	
	\item When the RTT of the current path reaches the threshold \texttt{th\_prob}, set to 12~$\mu$s in this example, the path probing module begins probing two alternative paths ($\text{P}_1$ and $\text{P}_4$) by creating two sets of QPs and binding them to different source ports (see Fig.~\ref{fig:hopper:probing}).
	
	\item If the RTT of the path exceeds the threshold \texttt{th\_cong}, set to 14~$\mu$s in this example, the path switching module compares the congestion on the current path ($\text{P}_3$) with that of the probed alternatives ($\text{P}_1$ and $\text{P}_4$). If a considerably better alternative path is available, the flow is switched to it after a cautious delay proportional to the delay difference between the current path ($\text{P}_3$) and the alternative path ($\text{P}_1$), to minimize OOO packets at the receiver NIC (see Fig.~\ref{fig:hopper:delaying}).
	
	\item Finally, the flow is switched to the selected alternative path ($\text{P}_1$) (see Fig.~\ref{fig:hopper:switching}).
\end{itemize}

\section{AliCloud Storage Workload}
\label{s:alibaba}
\begin{figure*}[t]
	\centering
	\begin{subfigure}[t]{0.45\textwidth}
		\centering
		\includegraphics[width=\textwidth]{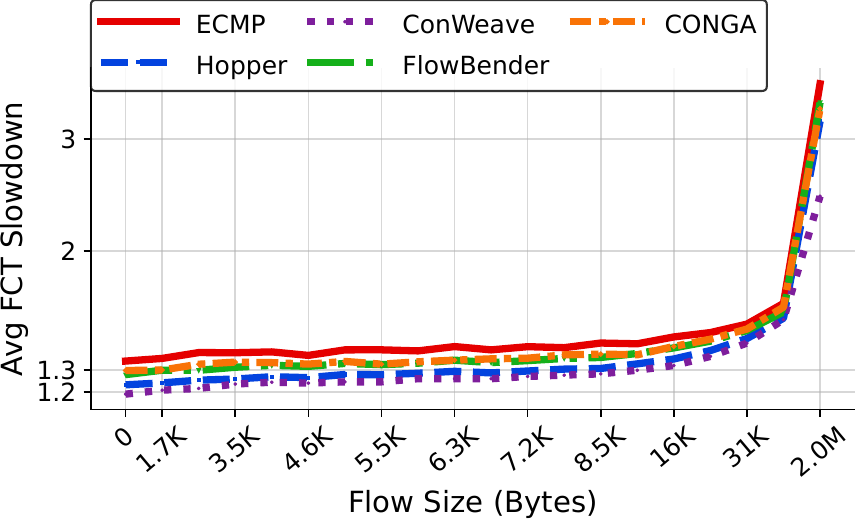}
		\caption{\scriptsize AliCloud 50\% Load (Avg)}
		\label{fig:alibaba:avg1}
	\end{subfigure}
	\hfill
	\begin{subfigure}[t]{0.45\textwidth}
		\centering
		\includegraphics[width=\textwidth]{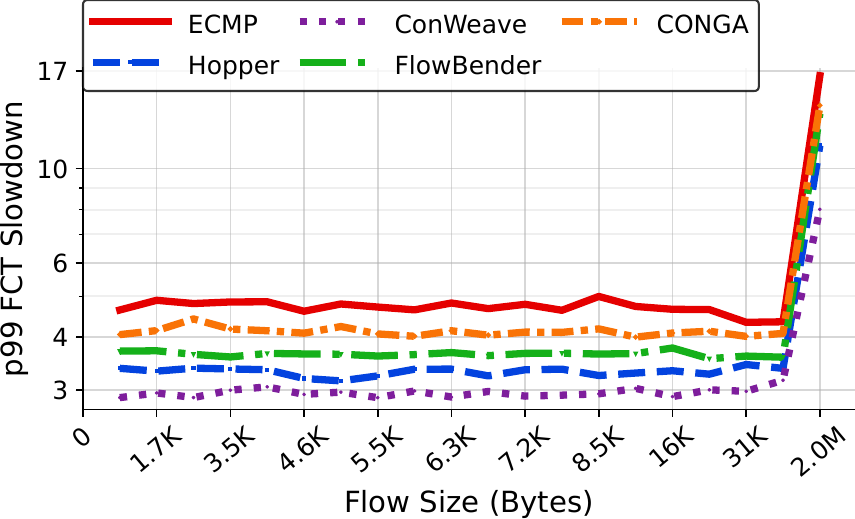}
		\caption{\scriptsize AliCloud 50\% Load (99\textsuperscript{th})}
		\label{fig:alibaba:p991}
	\end{subfigure}
	\begin{subfigure}[t]{0.45\textwidth}
		\vspace{2ex}		
		\centering
		\includegraphics[width=\textwidth]{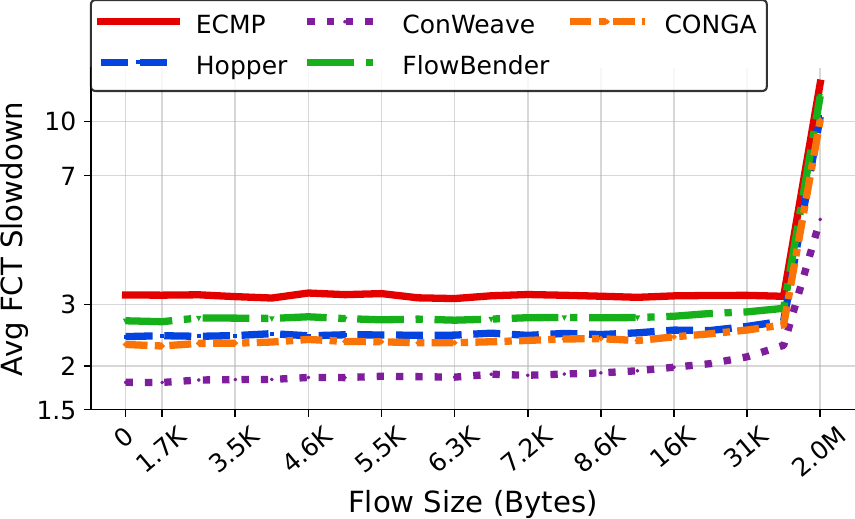}
		\caption{\scriptsize AliCloud 80\% Load (Avg)}
		\label{fig:alibaba:avg2}
	\end{subfigure}
	\hfill
	\begin{subfigure}[t]{0.45\textwidth}
		\vspace{2ex}		
		\centering
		\includegraphics[width=\textwidth]{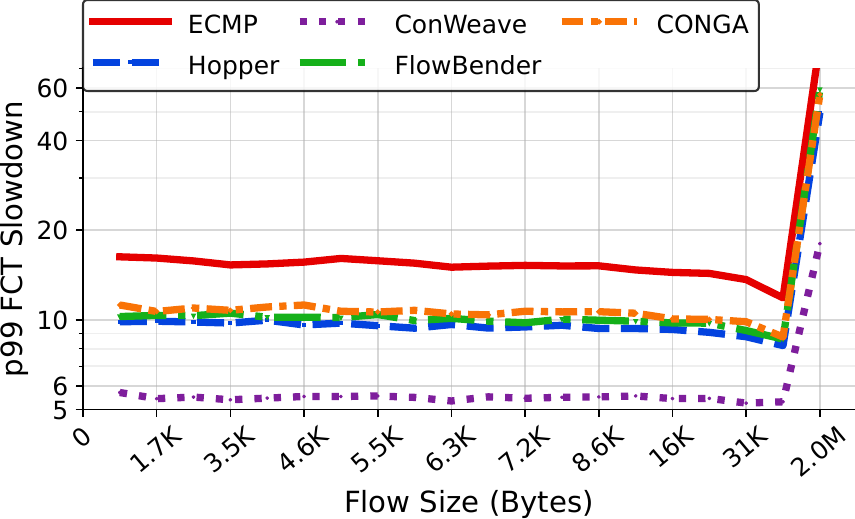}
		\caption{\scriptsize AliCloud 80\% Load (99\textsuperscript{th})}
		\label{fig:alibaba:p992}
	\end{subfigure}
	
	\caption{Average and tail FCT slowdown for AliCloud workload at $50\%$ and $80\%$ average network load.}
	\label{fig:alicloud:slowdowns}
	\vspace{2ex}		
\end{figure*}

We present the results of ns-3 simulation for AliCloud~\cite{HPCC} in Fig.~\ref{fig:alicloud:slowdowns} with the same parameters as for Meta Hadoop~\cite{Meta-hadoop}. Under 50\% load, our method provides up to \textbf{6\%} improvement for smaller flows and up to \textbf{19\%} for large flows on average, compared to CONGA and FlowBender. While it performs up to \textbf{6\%} worse than ConWeave for small flows, it still achieves up to \textbf{8\%} better 99th percentile performance than FlowBender and up to \textbf{16\%} better than CONGA. Under 80\% load, our approach continues to outperform FlowBender by approximately \textbf{10\%} in average, while showing similar behaviour for the tail latencies.

%
%

\section{Meta Hadoop Workload}
\begin{figure*}[t]
	\centering
	\begin{subfigure}[t]{0.45\textwidth}
		\centering
		\includegraphics[width=\textwidth]{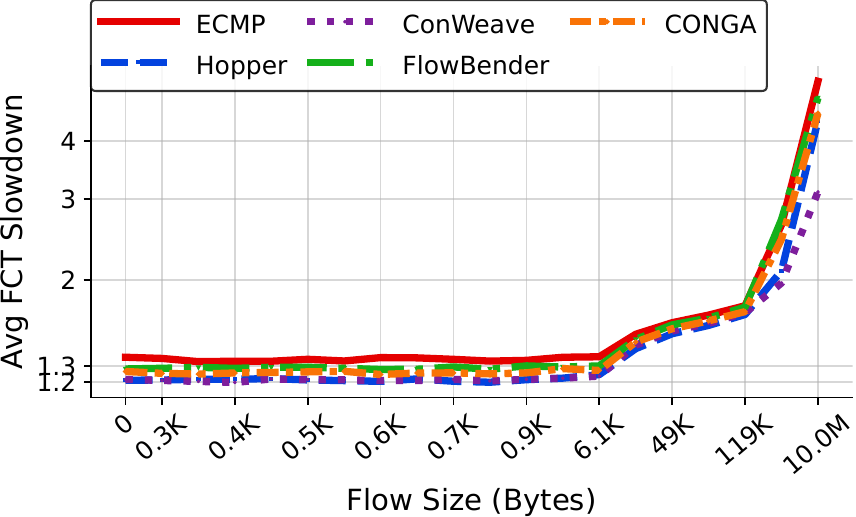}
		\caption{\scriptsize Hadoop 50\% Load (Avg)}
		\label{fig:Hadoop:avg1_complete}
	\end{subfigure}
	\hfill
	\begin{subfigure}[t]{0.45\textwidth}
		\centering
		\includegraphics[width=\textwidth]{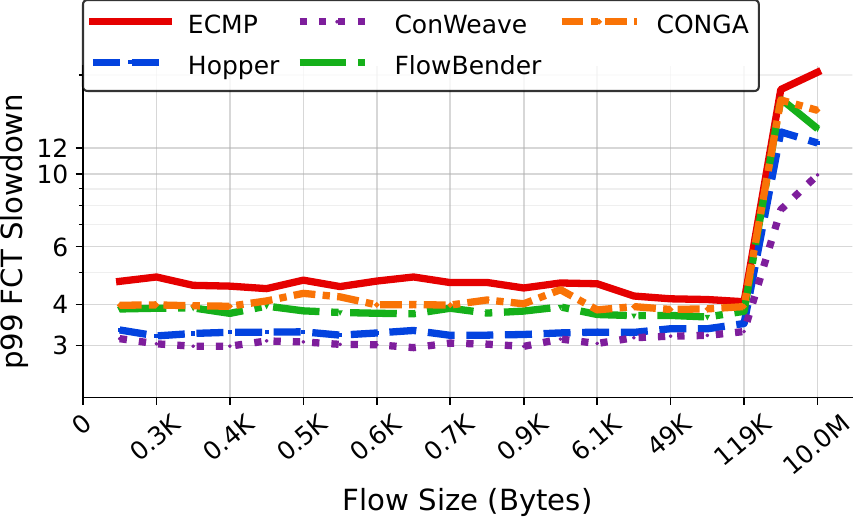}
		\caption{\scriptsize Hadoop 50\% Load (99\textsuperscript{th})}
		\label{fig:Hadoop:p991_complete}
	\end{subfigure}
	\begin{subfigure}[t]{0.45\textwidth}
		\vspace{2ex}		
		\centering
		\includegraphics[width=\textwidth]{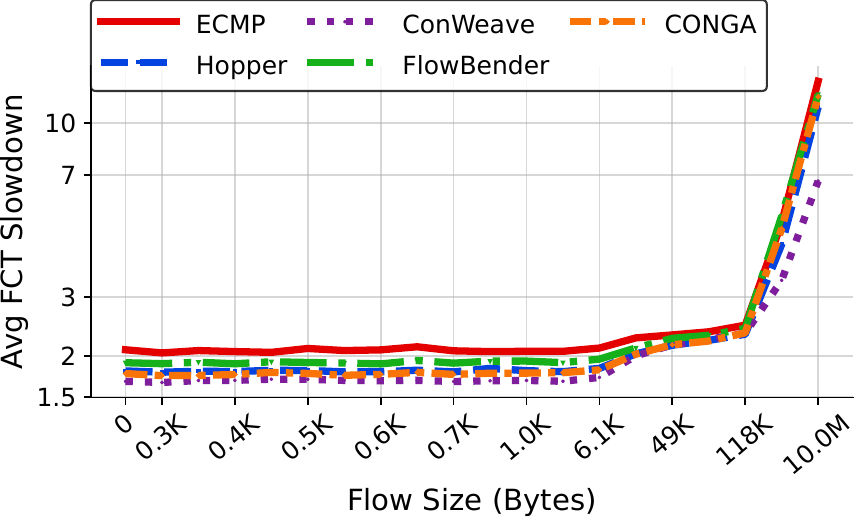}
		\caption{\scriptsize Hadoop 80\% Load (Avg)}
		\label{fig:Hadoop:avg2_complete}
	\end{subfigure}
	\hfill
	\begin{subfigure}[t]{0.45\textwidth}
		\vspace{2ex}		
		\centering
		\includegraphics[width=\textwidth]{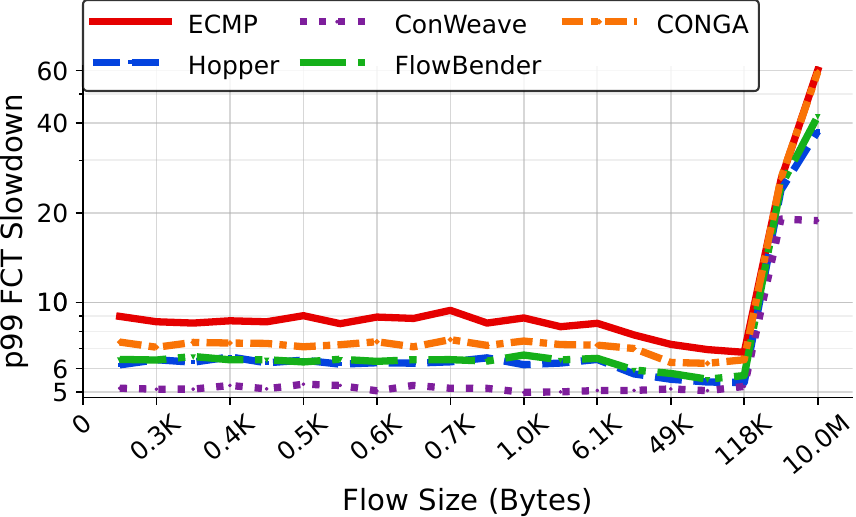}
		\caption{\scriptsize Hadoop 80\% Load (99\textsuperscript{th})}
		\label{fig:Hadoop:p992_complete}
	\end{subfigure}
	
	\caption{Average and tail FCT slowdown for Hadoop workload at $50\%$ and $80\%$ average network load.}
	\label{fig:Hadoop:slowdowns_complete}
\end{figure*}

For completeness, we include the full flow size distribution in Fig.\ref{fig:Hadoop:slowdowns_complete}. which covers all flow sizes including those between 2KB and 49KB which were omitted in the section~\ref{result_and_discussion} for clarity. 

\end{document}